\documentclass[aps,pra,amsmath,amssymb,floatfix,twocolumn,amsmath,superscriptaddress,twocolumn,nofootinbib,tighten,letterpaper]{revtex4-2}
\usepackage[colorlinks,linkcolor=blue,citecolor=blue,urlcolor=blue]{hyperref}
\usepackage{multirow}
\usepackage{subfigure}
\usepackage{color}
\usepackage{mathrsfs}
\usepackage{hyperref}
\usepackage[normalem]{ulem}
\usepackage{bm}

\usepackage{amssymb}   
\usepackage{amsmath}
\renewcommand\vec[1]{\ensuremath\boldsymbol{#1}} 

\usepackage{amsfonts, relsize, color}
\usepackage{graphics}
\usepackage{graphicx}
\usepackage{subfigure}
\usepackage{hyperref}
\usepackage{color}
\usepackage{comment}

\usepackage{array}

\newcolumntype{P}[1]{>{\centering\arraybackslash}p{#1}}

\usepackage{booktabs}

\begin{document}

\title{Two-dimensional Disordered Projected Branes: Stability and Quantum Criticality via Dimensional Reduction}

\author{Alexander C. Tyner}
\affiliation{Nordita, KTH Royal Institute of Technology and Stockholm University, Hannes Alfv\'ens v\"ag 12, 106 91 Stockholm, Sweden}
\affiliation{Department of Physics, University of Connecticut, Storrs, Connecticut 06269, USA}

\author{Vladimir Juri\v ci\' c}
\affiliation{Departamento de F\'isica, Universidad T\'ecnica Federico Santa Mar\'ia, Casilla 110, Valpara\'iso, Chile}

\author{Bitan Roy}~\thanks{Corresponding author: bitan.roy@lehigh.edu}
\affiliation{Department of Physics, Lehigh University, Bethlehem, Pennsylvania, 18015, USA}
\affiliation{Centre for Condensed Matter Theory, Physics Department, Indian Institute of Science, Bengaluru 560012, India}

\date{\today}

\begin{abstract}
The interplay of disorder and dimensionality governs the emergence and stability of electronic phases in quantum materials and quantum phase transitions among them. While three-dimensional (3D) dirty Fermi liquids and Weyl semimetals support robust metallic states, undergoing disorder-driven Anderson localization transitions at strong disorder and the later ones exhibiting additional semimetal-to-metal transition at moderate disorder, conventional two-dimensional (2D) non-interacting systems localize for arbitrarily weak disorder. Here, we show that 2D disordered projected branes, constructed by systematically integrating out degrees of freedom from a 3D cubic lattice via the Schur decomposition, faithfully reproduce the full quantum phase diagram of their 3D parent systems. Using large-scale exact diagonalization and kernel polynomial method, we numerically demonstrate that 2D projected branes host stable metallic and semimetallic phases. Remarkably, the critical exponents governing the semimetal-to-metal and metal-insulator transitions on such 2D projected branes are sufficiently close to those of their 3D
counterparts. Our findings thus establish 2D projected branes as genuine quantum holographic images of their higher-dimensional disordered parent crystals, supporting stable semimetallic and metallic phases that are otherwise inaccessible in conventional 2D lattices. Finally, we point to experimentally accessible metamaterial platforms, most notably the photonic lattices with tunable refractive-index disorder, as promising systems to realize and probe these phenomena.
\end{abstract}

\maketitle

\section{Introduction}

Can a $q$-dimensional subsystem, embedded within a $d$-dimensional electronic crystal, hereafter named \emph{brane} with $q < d$~\cite{panigrahi2022projected}, capture the topological properties and quantum phenomena, including disorder- and interaction-driven quantum phases, together with the critical behavior characterizing its higher-dimensional parent systems? Resolving this broad question is not only of fundamental theoretical interest, but also directly relevant for the design of quantum simulators and engineered metamaterials that can realize and probe emergent quantum and geometric phenomena arising via dimensional descent in controlled setups.


\begin{figure}[b!]
\includegraphics[width=1.00\linewidth]{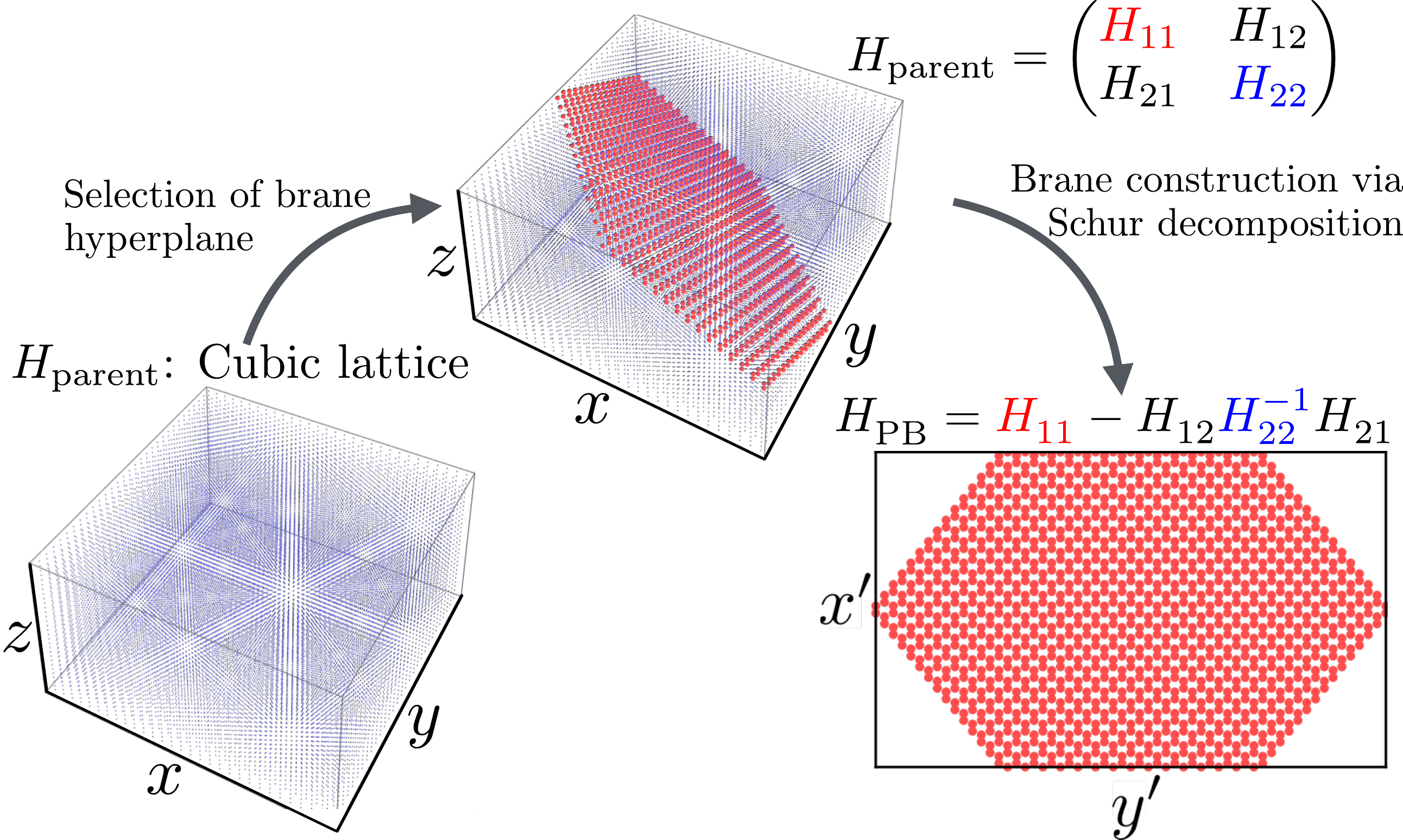}
\caption{An exemplary construction of a (hexagonal) \emph{projected brane} (PB) from a three-dimensional parent cubic lattice. Sites projected onto the brane are marked in red and the portion of the total Hamiltonian for the parent system $H_{\rm parent}$ exclusively operative among these sites corresponds to $H_{11}$ in Eqs.~\eqref{eq:parent} and~\eqref{eq:PTB}. Portion of $H_{\rm parent}$ that is exclusively operative among the sites shown in blue (falling outside the brane) yields $H_{22}$, and $H_{12}$ and $H_{21}(=H^\dagger_{12})$ results from the coupling between these two sets of sites. See Sec.~\ref{subsec:constructionPB} for the derivation of $H_{\rm PB}$, the effective Hamiltonian for PBs.   
}~\label{fig:construction}
\end{figure}

As a concrete example, we consider the case of a two-dimensional (2D) hexagonal brane embedded within a three-dimensional (3D) parent cubic lattice, as depicted in Fig.~\ref{fig:construction}. By constructing an effective (quadratic or non-interacting) Hamiltonian for such a hexagonal brane upon projecting the parent cubic lattice Hamiltonian following the Schur decomposition~\cite{panigrahi2022projected, tynerjuricic2024, DJSalib2024, Schur1917} (detailed in Sec.~\ref{subsec:constructionPB}), here we set out to answer the following question. Can such a 2D brane capture the electronic properties of disordered (due to random pointlike charge impurities) 3D parent cubic lattice? Our extensive numerical analysis establishes that the 2D projected branes (PBs) not only support stable ballistic and diffusive phases of non-interacting fermions but also exhibit disorder-driven metal-insulator and semimetal-metal quantum phase transitions, with critical exponents quantitatively matching those of the 3D parent systems (within the numerical accuracy). In turn, this outcome provides a direct route to stabilize dirty 2D disordered Fermi and Weyl liquids, thereby bypassing the requirement of strong interactions for planar non-Fermi liquids to remain robust against weak disorder~\cite{ChakravartyAbrahams1998}. For concreteness, we focus on the vastly studied lattice-regularized Anderson~\cite{Anderson1958, thouless1970, AALR1979, Wegner1979, Lee1981, Hikami1981, RevModPhys1985AT, Kirkpatrick1994, janssen1998, slevin1999, slevin2001, klopp2002, dobrosavljevic2003, Brndiar2006, Chakravarty2008, RevModPhys2008AT, slevin2010} and Weyl~\cite{Fradkin1986, Murakami2009, goswami2011, Imura2013, Herbut2014, roydassarma2014, Syzranov2015a, Syzranov2015b, Altland2015, pix2015, Syzranov2016, Pix2016a, roydassarma2016, bera2016, Gorbar2016, RoyJuricicDasSarma2016, RoyDasSarma2016PRB, Carpentier2016, Carpentier2017, Justin2017, Goswami2017a, Goswami2017b, Alavirad2017, Slager2017, WilsonRefael2018, Slager2018, Carpentier2018, Carpentier2019, Szabo2020, Gruzberg2024} models in three dimensions, constituting the set of parent systems, subject to random pointlike charge impurities and study their imprints on the global phase diagram of 2D hexagonal PBs. To set the stage, we summarize the effects of disorder on electronic fluids in three and two dimensions.

\subsection{Dirty electronic liquids: A brief review}

To proceed, it is useful to classify electronic fluids according to the scaling of the density of states (DOS) in clean systems. This approach divides the landscape into two broad classes: conventional Fermi liquids, which exhibit a finite DOS at the band center, and nodal Fermi liquids, which feature a DOS that vanishes as a power law at the band center. Throughout this work, we consider models with particle-hole symmetric spectra that extend to both positive and negative energies, such that the band center is located at $E = 0$ in the clean limit.

\begin{figure}
\includegraphics[width=1.00\linewidth]{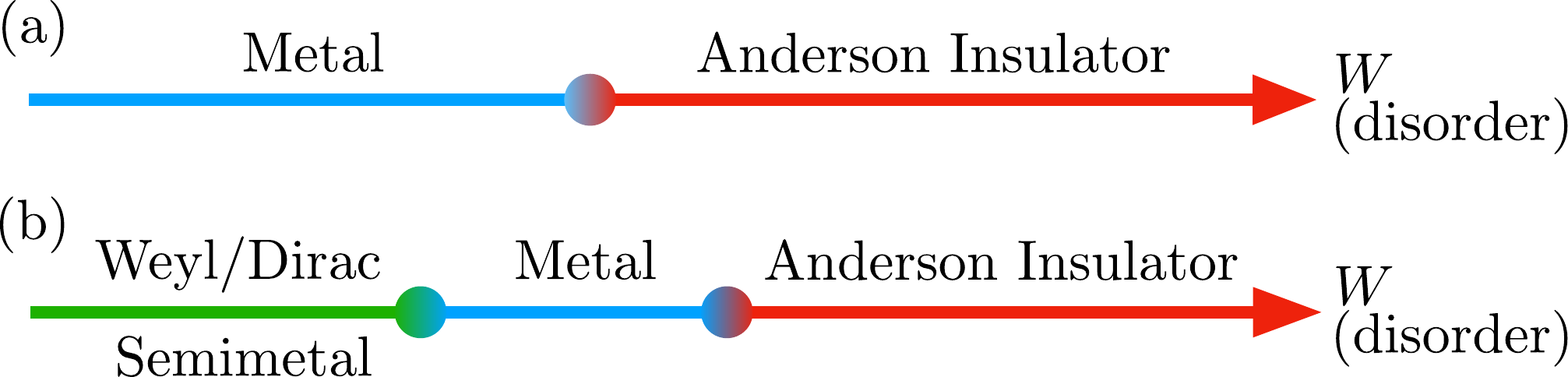}
     \caption{(a) Schematic phase diagram of a three-dimensional (3D) disordered Fermi liquid or Anderson model and a two-dimensional (2D) projected brane constructed from parent 3D Anderson model on a cubic lattice. (b) Schematic phase diagram of a disordered 3D Weyl/Dirac liquid and a 2D projected Weyl brane constructed from the 3D Weyl/Dirac semimetal on a cubic lattice. For the construction of the 2D projected branes from the parent cubic lattices see Fig.~\ref{fig:construction} and Sec.~\ref{subsec:constructionPB}. Circles separating two adjacent phases represent the quantum critical points, controlling the continuous phase transition between them. In two dimensions neither a metal nor the Dirac or Weyl semimetal is a stable phase of matter even for infinitesimal strength of disorder. Systems then always exist in the Anderson insulator phase in the thermodynamic limit. The disorder strength due to random pointlike charge impurities is denoted by $W$. 
     }~\label{fig:Phasediag}
\end{figure}

The impact of disorder on conventional Fermi liquids has been the subject of intense investigation over the past several decades, with Anderson localization and the associated metal-insulator quantum phase transition in three dimensions ($d=3$) occupying a central stage~\cite{Anderson1958, thouless1970, AALR1979, Wegner1979, Lee1981, Hikami1981, RevModPhys1985AT, Kirkpatrick1994, janssen1998, slevin1999, slevin2001, klopp2002, dobrosavljevic2003, Brndiar2006, Chakravarty2008, RevModPhys2008AT, slevin2010}. In three dimensions, disorder must exceed a critical strength to localize all the electronic wave functions, thereby producing an Anderson insulating state, as schematically illustrated in Fig.~\ref{fig:Phasediag}(a). By contrast, no such Anderson transition exists in dimensions $d \leq 2$ and Fermi liquids in two dimensions ultimately localize for arbitrarily weak disorder, although the localization occurs logarithmically slowly with increasing system size~\cite{AALR1979}, leaving aside sparse exceptions such as 2D symplectic class models~\cite{Wegner1989, Evangelou1987, Ando1989PRB, Evangelou1995, Asada2002, Ando2004} and hyperbolic lattices~\cite{Boettcher2024, TianyuLi2024} that also display Anderson metal-insulator transition. These findings underscore the pivotal role of dimensionality in determining the low-temperature behavior of dirty electronic systems, which becomes even more prominent for certain nodal Fermi liquids, as discussed below.

\begin{table*}[t!]
\centering
\begin{tabular}{
|P{1.75cm}|P{2cm}|
P{1.5cm}|P{3.2cm}|P{1.5cm}|
P{1.2cm}|P{3cm}|P{1.2cm}|}
\hline
{\vspace{.3cm}\textbf{Model}} & {\vspace{.3cm}\textbf{Transition}} & 
\multicolumn{3}{c|}{\raisebox{-0.3cm}{\textbf{Parent cubic lattice}}} & 
\multicolumn{3}{c|}{\raisebox{-0.3cm}{\textbf{2D Projected brane}}} \\
\cline{3-8}
& & Method & Result & Figure & Method & Result & Figure \\
\hline\hline
\multirow{2}{*}{\raisebox{-0.55cm} {Anderson}} & \multirow{2}{*} {\raisebox{-0.55cm}{MIT}} 
& {{\raisebox{0.15cm} {KPM}} {\raisebox{0.15cm} {(TDOS)}}} & 
\parbox[c][.8cm][b]{3.2cm}{\centering 
$W_c=3.10\pm0.10$\\[1pt]$\beta=1.60\pm0.16$} & Fig.~\ref{fig:AndersonKPM} 
& {{\raisebox{0.15cm} {ED$_1$}} {\raisebox{0.15cm} {(TDOS)}}} & 
\parbox[c][.8cm][b]{3.2cm}{\centering 
$W_c = 17.95 \pm 0.50$ \\ $\beta = 1.55 \pm 0.10$} & Fig.~\ref{fig:BraneAnderson} \\
\cline{3-8}
& & {{\raisebox{0.15cm} {ED}} {\raisebox{0.15cm} {(TDOS)}}} & 
\parbox[c][.8cm][b]{3.2cm}{\centering 
$W_c = 3.50 \pm 0.15$\\ $\beta=1.60\pm0.14$} & Fig.~\ref{fig:BraneAnderson} 
& {{\raisebox{0.15cm} {ED$_2$}} {\raisebox{0.15cm} {(TDOS)}}} & 
\makebox[3.0cm][c]{-----} & \makebox[1.2cm][c]{-----} \\ 
\hline\hline
\multirow{6}{*}{\raisebox{-2.4cm} {Dirty Weyl}} & \multirow{4}{*}{\raisebox{-1.5cm} {SMMT}} 
& {{\raisebox{0.25cm} {KPM}} {\raisebox{0.25cm} {(ADOS)}}} & 
\parbox[c][1.1cm][b]{3.2cm}{\centering 
$W_c = 0.50 \pm 0.05$\\$\alpha=1.00 \pm 0.05$\\ $\beta = 1.55 \pm 0.10$} & Fig.~\ref{fig:WeylKPM}
& {{\raisebox{0.25cm} {ED$_1$}} {\raisebox{0.25cm} {(ADOS)}}} & 
\parbox[c][1.1cm][b]{3.0cm}{\centering 
$W_c=0.70\pm 0.02$\\ $\alpha= 1.01 \pm 0.05$\\ $\beta=1.55\pm 0.10$} & Figs.~\ref{fig:PBWeylADOS} and~\ref{fig:exponents} \\
\cline{3-8}
& & {{\raisebox{0.25cm} {KPM}} {\raisebox{0.25cm} {(TDOS)}}} & 
\parbox[c][.8cm][b]{3.2cm}{\centering 
$W_c=0.70\pm 0.10$\\ $\beta=1.75\pm 0.15$} & Fig.~\ref{fig:WeylKPMTDOS}
& {{\raisebox{0.25cm} {ED$_2$}} {\raisebox{0.25cm} {(ADOS)}}} & 
\parbox[c][1.1cm][b]{3.0cm}{\centering 
$W_c=0.70\pm 0.02$\\ $\alpha= 0.98 \pm 0.05$\\ $\beta=1.50\pm 0.05$} & Figs.~\ref{fig:PBWeylADOS} and~\ref{fig:exponents} \\
\cline{3-8}
& & {{\raisebox{0.25cm} {ED}} {\raisebox{0.25cm} {(ADOS)}}} & 
\parbox[c][1.2cm][b]{3.2cm}{\centering 
$W_c = 0.49 \pm 0.05$\\ $\alpha=1.00\pm0.03$ \\$\beta=1.50 \pm 0.05$} & Fig.~\ref{fig:exponents} 
& {{\raisebox{0.25cm} {ED$_1$}} {\raisebox{0.25cm} {(TDOS)}}} & 
\parbox[c][.8cm][b]{3.0cm}{\centering 
$W_c=0.75\pm0.10$\\ $\beta=1.80 \pm 0.10$} & Fig.~\ref{fig:WeylAnderson} \\
\cline{3-8}
& & ED (TDOS) & 
\raisebox{-0.15cm}{\makebox[3.2cm][c]{-----}} & 
\raisebox{-0.15cm}{\makebox[1.5cm][c]{-----}} 
& ED$_2$ (TDOS) & 
\raisebox{-0.15cm}{\makebox[3.0cm][c]{-----}} & 
\raisebox{-0.15cm}{\makebox[1.2cm][c]{-----}} \\
\cline{2-8}
& \multirow{2}{*}{\raisebox{-.5cm} {MIT}} 
& {{\raisebox{0.2cm} {KPM}} {\raisebox{0.2cm} {(TDOS)}}} & 
\parbox[c][.8cm][b]{3.2cm}{\centering 
$W_c=3.85\pm 0.15$\\ $\beta=1.59\pm 0.11$} & Fig.~\ref{fig:WeylKPMTDOS} 
& {{\raisebox{0.2cm} {ED$_1$}} {\raisebox{0.2cm} {(TDOS)}}} & 
\parbox[c][.8cm][b]{3.0cm}{\centering 
$W_c=20.00\pm 1.00$\\ $\beta=1.60\pm 0.15$} & Fig.~\ref{fig:WeylAnderson} \\
\cline{3-8}
& & {{\raisebox{0.01cm} {ED}} {\raisebox{0.01cm} {(TDOS)}}} & 
\raisebox{-0.15cm}{\makebox[3.2cm][c]{-----}} & 
\raisebox{-0.15cm}{\makebox[1.5cm][c]{-----}} 
& {{\raisebox{0.01cm} {ED$_2$}} {\raisebox{0.01cm} {(TDOS)}}} & 
\raisebox{-0.15cm}{\makebox[3.0cm][c]{-----}} & 
\raisebox{-0.15cm}{\makebox[1.2cm][c]{{-----}}} \\
\hline
\end{tabular}
\caption{Summary of the results for the three-dimensional (3D) Anderson model and Dirty Weyl model, as well as for the corresponding two-dimensional (2D) projected branes (PBs). While the former system displays only Anderson metal-insulator transition (MIT), the dirty Weyl system in addition displays semimetal-to-metal transition (SMMT) in three dimensions. These transitions are also observed on their 2D PBs. While the MIT and SMMT is identified from the scaling of the typical density of states (TDOS) at zero energy, the average density of states (ADOS) at zero energy emerges as an additional order parameter across the SMMT. These quantities are computed using the kernel polynomial method (KPM) on larger cubic lattices and the exact diagonalization (ED) in smaller cubic lattices. On 2D PBs, TDOS and ADOS are always computed using the ED. But, we employ two methods in that calculation by projecting the clean cubic lattice models onto the branes and subsequently adding disorder to them (ED$_1$) or by directly projecting the disordered 3D Hamiltonian onto the branes (ED$_2$). Here, $W_c$ is the critical disorder strength for the corresponding transition. While $\alpha$ is the critical exponent associated with the scaling of the density of states with energy near the SMMT, $\beta$ is the order parameter exponent associated with ADOS and TDOS near the SMMT and MIT transitions, but within the metallic phase. We also refer to the corresponding figure for each case.}
\label{tab:summary_results}
\end{table*}

The global phase diagram of disordered nodal Fermi liquids can be dramatically distinct from that of conventional dirty Fermi liquids, particularly in three dimensions. Specifically, if the DOS near the band center scales as $\rho(E) \sim |E|^n$, a generalized Harris criterion can be invoked to assess the stability of such nodal liquids against infinitesimal disorder~\cite{Slager2018, Harris1974, CCFS1986}. This power-law scaling of the DOS typically arises from the touching of the filled valence and empty conduction bands at isolated points or nodes in the Brillouin zone, so-called nodal points, giving rise to the class of electronic fluids known as nodal Fermi liquids. In this context, ballistic quasiparticles in a nodal Fermi liquid are stable against sufficiently weak disorder when $n > 1$~\cite{Slager2018}. In $d=3$, this criterion is readily satisfied for Dirac and Weyl liquids, which feature a linear energy-momentum dispersion in all directions, leading to $\rho(E) \sim |E|^2$. Consequently, Dirac and Weyl liquids display two \emph{distinct} quantum phase transitions as disorder strength increases. First, they undergo a continuous transition into a diffusive metallic phase, where $\rho(0)$ becomes finite; subsequently, the resulting Weyl metal can experience a conventional Anderson localization transition at even stronger disorder, as shown schematically in Fig.~\ref{fig:Phasediag}(b). The  semimetal-metal transition, unique to Dirac and Weyl systems and fundamentally distinct from the extensively studied Anderson transition, has triggered a new wave of investigations about a decade ago, geared toward unfolding the nature of such semimetal-to-metal quantum phase transition~\cite{Fradkin1986, Murakami2009, goswami2011, Imura2013, Herbut2014, roydassarma2014, Syzranov2015a, Syzranov2015b, Altland2015, pix2015, Syzranov2016, Pix2016a, roydassarma2016, bera2016, Gorbar2016, RoyJuricicDasSarma2016, RoyDasSarma2016PRB, Carpentier2016, Carpentier2017, Justin2017, Goswami2017a, Goswami2017b, Alavirad2017, Slager2017, WilsonRefael2018, Slager2018, Carpentier2018, Carpentier2019, Szabo2020, Gruzberg2024}.

In two spatial dimensions, however, Dirac and Weyl systems are not stable against even sufficiently weak disorder. Specifically, infinitesimal disorder converts these systems into a putative diffusive metal, which manifests in a crossover regime and ultimately evolves into an Anderson insulator in the thermodynamic limit. It is important to note that this outcome strictly applies when the system hosts an even number of Dirac or Weyl nodes, as required by the Nielsen-Ninomiya no-go theorem for lattice-regularized models under rather general conditions~\cite{Nielsen1981a, Nielsen1981b}. By contrast, three-dimensional strong $\mathbb{Z}_2$ topological insulators can host an odd number of Weyl nodes on their 2D surfaces~\cite{Ludwig2008, Ludwig2010}. Owing to the absence of backscattering, such surface Weyl fermions exhibit supermetallic behavior for infinitesimal disorder, devoid of any localization transition. This is so because the corresponding non-linear $\sigma$ model is accompanied by a topological $\theta$-term with a nontrivial $\theta=\pi$ (modulo $2\pi$), allowing two-component planar Weyl or Dirac fermions to evade the Anderson localization and constitute a supermetal phase for arbitrarily strong disorder~\cite{Mirlin2007PRL, Bardarson2007PRL, Furusaki2007PRL, Ryu2007PRL}. Here,  we do not delve into this situation.

As a penultimate point in this section, it should be noted that the sharpness of the Dirac or Weyl semimetal-to-metal quantum phase transition in $d=3$ can, in principle, be masked by \emph{non-perturbative} rare region effects~\cite{Huse2014, Huse2016a, Huse2016b, buch2018, JustinWilson2020, JustinWilson2024, PixleyWilsonReview2021}. Although rare region effects and the associated Griffiths phase are believed to set in at infinitesimal disorder strength, their signatures in numerical simulations have thus far only been observed in close proximity to the semimetal-to-metal quantum critical point, accessible via perturbative field-theoretic approaches. In this work, we do not address rare region effects on 2D PBs.

As a final comment, we discuss the order parameters that characterize the Anderson metal-insulator and Dirac or Weyl semimetal-to-metal quantum phase transitions. The typical density of states (TDOS) at the band center, $\rho_{\rm typ}(E=0) \equiv \rho_{\rm typ}(0)$, serves as a bona fide order parameter for both transitions; it is finite in the diffusive metallic phase, but vanishes in the Anderson insulating and Dirac/Weyl semimetal phases. In contrast, the average density of states (ADOS) at zero energy, $\rho_{\rm av}(E=0) \equiv \rho_{\rm av}(0)$, plays a role of an order parameter for the semimetal-to-metal transition only; it vanishes in the semimetal and becomes finite in the metal, but remains noncritical across the Anderson transition. Accordingly, we numerically compute both $\rho_{\rm typ}(0)$ and $\rho_{\rm av}(0)$ to construct the global phase diagrams of 3D disordered conventional Fermi and Weyl liquids, as well as their projected counterparts on 2D branes. In this way, we establish a one-to-one correspondence between the global phase diagrams of the disordered parent quantum crystals and their PB counterparts, which is the central theme of this work. The central results of the current work are summarized in Table~\ref{tab:summary_results}. Next we outline the procedure to construct PBs.

\subsection{Projected branes: Construction}~\label{subsec:constructionPB}

A systematic algorithm for constructing PBs was introduced in Ref.~\cite{panigrahi2022projected}, subsequently generalized to arbitrary dimensions in Ref.~\cite{tynerjuricic2024}, and recently extended to regular fractal lattices in Ref.~\cite{DJSalib2024}. We briefly summarize this procedure here to set the stage for our central results, presented in Sec.~\ref{subsec:mainresults}. The construction of the PB Hamiltonian is based on the Schur complement~\cite{Schur1917}. Consider a quadratic tight-binding Hamiltonian defined on a $d$-dimensional hypercubic lattice with $L^d$ sites, referred to as the parent system. In real space, the parent Hamiltonian can be written in block matrix form as
\begin{equation}~\label{eq:parent}
    H_{\rm parent}=\left( \begin{array}{cc}
    H_{11} & H_{12} \\
    H_{21} & H_{22}
    \end{array} \right),
\end{equation} 
where $H_{11}$ corresponds to the real-space Hamiltonian for the lattice sites on the $q$-dimensional PB, while $H_{22}$ refers to the real-space Hamiltonian for the subsystem consisting of sites outside the PB. The off-diagonal blocks, $H_{12}$ and $H_{21} = H_{12}^\dagger$, capture the coupling between the PB and the rest of the parent lattice. This construction is illustrated in Fig.~\ref{fig:construction} for a two-dimensional hexagonal PB, formed from a plane perpendicular to the body diagonal of a cubic lattice. The Hamiltonian for the PB is then obtained by systematically integrating out the sites that fall outside the PB and it takes the form~\cite{panigrahi2022projected}
\begin{equation}\label{eq:PTB}
H_{\rm PB}(\kappa) = H_{11} - \kappa \left( H_{12} H_{22}^{-1} H_{21} \right).
\end{equation}
Although this approach can be applied to any $q$-dimensional PB with $q < d$, here we focus on PBs with $q = d-1$ and $d=3$. In Eq.~\eqref{eq:PTB}, the parameter $\kappa$ is introduced for generality. Unless otherwise specified, we set $\kappa=1$ (yielding the exact Schur complement result), whereas setting $\kappa=0$ we can assess the impact of integrating out the degrees of freedom falling outside the PB, captured by the second term. Unless specified, we use the notation $H_{\rm PB}(\kappa=1) \equiv H_{\rm PB}$.

The lattice sites which constitute the PB are those that lie within a specified distance (set by the lattice spacing $a$) from a $(d-1)$-dimensional hyperplane defined within a $d$-dimensional parent system. The equation of such a hyperplane takes the generic form 
\begin{equation}~\label{eq:hyper}
    \sum_{j=1}^{d} \gamma_{j}x_{j}=\eta,
\end{equation}
where $\gamma_{j}$ and $\eta$ are real numbers. A lattice site $i$, identified by its coordinates $\{ x_j (i) \}$, projected onto the $(d-1)$-dimensional brane, obeys the relation~\cite{tynerjuricic2024}
\begin{equation}~\label{eq:const}
\left( \sqrt{\sum_{j=1}^{d}\gamma_{j}^2} \; \right)^{-1} \times  \left\vert \; \sum_{j=1}^{d}\gamma_{j} x_{j}(i)+\eta \; \right\vert
<\frac{1}{\sqrt{2} \; a}.
\end{equation}
Throughout we fix the parameters for our 2D hyperplane, $\alpha_{j}=1$ for $j=1,2,3$ and $\eta=1/100$ such that the brane is formed from sites projected on a hexagonal hyperplane perpendicular to the body diagonal shown in Fig.~\ref{fig:construction}. The specific choice of such an orientation of the PB within the parent cubic lattice is justified in the following way. All the parent cubic lattice-based model Hamiltonian we consider in this work (Anderson and Weyl) feature cubic symmetry in the clean limit (although the resulting states are not protected by any crystalline symmetry). Only the 2D hyperplane, perpendicular to one of the body diagonals of the cubic lattice, respects that symmetry. However, we believe that our conclusions are insensitive to the choice of the orientation of 2D hyperplanes, constituting the PBs, which we do address explicitly here. Nonetheless, this anticipation can be justified from a different context. A contemporary work has shown that discrete four-fold rotational ($C_4$) symmetry protected crystalline topological insulators, realized on a square lattice when projected onto a one-dimensional brane, featuring either Fibonacci quasi-crystal or its rational approximant with emergent periodicity, they retain all the salient crystalline topological properties despite such projected one-dimensional branes being devoid of the $C_4$ symmetry of the parent square lattice.

Altogether, PBs constitute a fascinating route to capture quantum-critical and topological properties of a $d$-dimensional system on its $q$-dimensional geometric descendants, namely the PBs. Prior works have demonstrated that $(d-1)$-dimensional PBs of $d$-dimensional parent topological insulators feature electronic topological properties of the parent system in $(d-1)$-dimensions, paving a realistic pathway to traverse through the Altland-Zirnbauer table to harness topological phases beyond three spatial dimensions~\cite{Ludwig2008, Ludwig2010} and possibly realize topological states beyond the ten-fold way~\cite{panigrahi2022projected, tynerjuricic2024}. For example, robust zero-energy chiral edge modes, hallmarks of a 2D class A quantum anomalous Hall insulator, can emerge at the endpoints of a one-dimensional chain, which can be either a Fibonacci quasicrystal or its rational approximant, protected by a quantized local Chern marker~\cite{panigrahi2022projected}. This observation is especially striking, given that class A systems in one dimension do not, by themselves, support any nontrivial topological phases according to the Altland-Zirnbauer classification.

In this work, we address the important question of whether PBs inherit the stability to disorder exhibited by their 3D parent Hamiltonian. An affirmative answer establishes PBs as genuine geometric descendants of their parent dirty quantum crystals, realized through the dimensional reduction. In this context, we now present a synopsis of our central results.

\subsection{Summary of main results}~\label{subsec:mainresults}

\begin{figure*}
\includegraphics[width=1.00\linewidth]{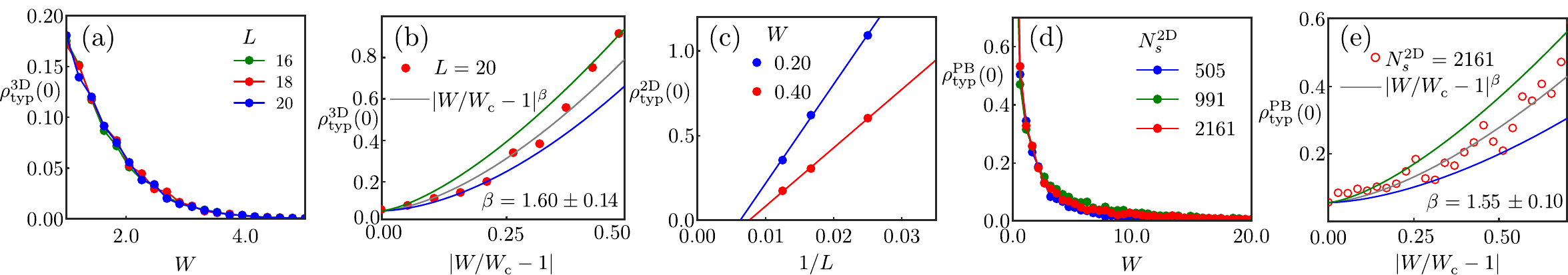}
     \caption{(a) Typical density of states (TDOS) at zero energy [see Eq.~\eqref{eq:tdos}], $\rho^{\rm 3D}_{\rm typ}(0)$, computed for the three-dimensional (3D) Anderson model on a cubic lattice [see Eq.~\eqref{eq:anderson}] of various linear dimension $L$ (mentioned in the legend) and periodic boundary condition (PBC) in each direction as a function of the disorder strength ($W$) after averaging over 500 independent disorder configurations from exact diagonalization (ED). Notice that $\rho^{\rm 3D}_{\rm typ}(0)$ becomes zero only beyond a critical strength of disorder $W_c=3.50 \pm 0.15$, marking the quantum critical point associated with the Anderson metal-insulator transition. (b) Computation of the order-parameter exponent $\beta$ across the Anderson transition on a cubic lattice of $L=20$, obtained from the scaling of TDOS at zero-energy in the metallic phase around $W_c$ [see Eq.~\eqref{eq:Andersonscaling}]. (c) A linear fit to the TDOS at zero-energy for two sufficiently small disorder strengths ($W=0.2$ and $0.4$) for a two-dimensional (2D) Anderson model on a square lattice of linear dimension $L$ and PBC in each direction as a function of $1/L$, denoted by $\rho^{\rm 2D}_{\rm typ}(0)$. It shows that even for such small disorder strengths $\rho^{\rm 2D}_{\rm typ}(0) \to 0$ before the system reaches the thermodynamic limit ($1/L \to 0$), indicating the absence of any stable metallic phase in two dimensions. Each data point is obtained from ED after averaging over 500 independent disorder realizations. (d) TDOS at zero energy for the projected brane (PB), denoted by $\rho^{\rm PB}_{\rm typ}(0)$, constructed from the 3D parent cubic lattice Anderson model (see Fig.~\ref{fig:construction} and Sec.~\ref{subsec:constructionPB}) after averaging over 500 independent disorder configurations, as a function of $W$ with varying number of sites $N^{\rm 2D}_s$ within such a PB (mentioned in the legend). Results are obtained from ED. Notice that the scaling of $\rho^{\rm PB}_{\rm typ}(0)$ with $W$ is qualitatively similar to that for $\rho^{\rm 3D}_{\rm typ}(0)$, suggesting that the PB also displays a metal-insulator transition at a finite disorder $W_c=17.95 \pm 0.50$, similar to the situation in the parent 3D cubic Anderson model; see panel (a). (e) Computation of the order-parameter exponent $\beta$ across the Anderson transition on a 2D PB, obtained by following the same procedure as in (b). Notice that the critical exponent $\beta$, setting universality class of the Anderson transition, assumes similar values on a cubic lattice and 2D PB (within numerical accuracy). Disorder-averaged TDOS is computed by averaging the disorder-averaged local density of states over $N_t=12$ [in (a)] and $4$ [in (c) and (d)] number of sites; see Eq.~\eqref{eq:LDOS}.      
     }~\label{fig:BraneAnderson}
\end{figure*}

Focusing first on the 3D Anderson model on a cubic lattice, we demonstrate that its geometric descendant, the 2D hexagonal PB, exhibits the same phase diagram as the parent system. Specifically, we find a stable metallic phase on the 2D PB up to a critical disorder strength, beyond which the system transitions to an Anderson insulator. The associated critical exponent is also in reasonable agreement with that of the 3D cubic lattice model~\cite{slevin2001}. By contrast, neither a conventional 2D square lattice Anderson model nor $H_{\rm PB}(0)$ from Eq.~\eqref{eq:PTB} displays a stable metallic phase. These observations confirm that the 2D PB inherits the phase diagram of the original cubic lattice Anderson model, as summarized in Fig.~\ref{fig:BraneAnderson} and Table~\ref{tab:summary_results}.

Next, we investigate Weyl liquids in both three and two dimensions. In finite systems, our model Hamiltonian for clean Weyl systems yields a finite number of zero-energy modes due to momentum quantization (see Appendix~\ref{append:zeromodes}). To correctly identify the semimetal-to-metal quantum phase transition, we subtract the contributions of these zero-energy modes, resulting in a vanishing DOS in the clean limit, as shown in Fig.~\ref{fig:DOSsubtraction}. Upon introducing weak disorder, these modes shift away from zero energy, and we recover the expected energy dependence of the DOS: quadratic scaling in $d=3$ and linear scaling in $d=2$, as displayed in Fig.~\ref{fig:DOSWeyl2D3D}. We first confirm that a Weyl Hamiltonian defined on a cubic (square) lattice in $d=3$ ($d=2$) exhibits (does not exhibit) a semimetal-to-metal quantum phase transition at finite disorder strength, consistent with established results~\cite{Fradkin1986, Murakami2009, goswami2011, Imura2013, Herbut2014, roydassarma2014, Syzranov2015a, Syzranov2015b, Altland2015, pix2015, Syzranov2016, Pix2016a, roydassarma2016, bera2016, Gorbar2016, RoyJuricicDasSarma2016, RoyDasSarma2016PRB, Carpentier2016, Carpentier2017, Justin2017, Goswami2017a, Goswami2017b, Alavirad2017, Slager2017, WilsonRefael2018, Slager2018, Carpentier2018, Carpentier2019, Szabo2020, Gruzberg2024}, as shown in Fig.~\ref{fig:DOSWeyl2D3D}, which here we find by employing  exact diagonalization. Having validated these features for conventional Weyl systems, we then investigate the effects of disorder on the projected hexagonal Weyl brane constructed from the parent cubic lattice, leading to the following key findings.

\begin{table}[t!]
\begin{tabular}{|c|c|c||c|c|}
\hline
$L$ & $N^{\rm 2D}_s$ & ${\rm dim}.(H^{3 \to 2}_{22})$ & $N^{\rm 1D}_s$ & ${\rm dim}.(H^{3 \to 1}_{22})$ \\
\hline \hline
11 & 271 & 1060 & 11 & 1320 \\
\hline
15 & 505 & 2870 & 15 & 3360 \\
\hline
21 & 991 &  8270 & 21 & 9240 \\
\hline
31 & 2161 & 27630 & 31 & 29760 \\
\hline
35 & 2755 & 40120 & 35 & 42840 \\
\hline
\end{tabular}
\caption{A few examples showing the linear dimension of the parent three-dimensional (3D) cubic lattice $L$ in each direction (first column), the number of sites contained within the two-dimensional (2D) projected brane $N^{\rm 2D}_s$ (second column), and dimensionality of the matrix $H_{22}$, here denoted by ${\rm dim}.(H^{3 \to 2}_{22})$ (third column), that needs to be inverted to obtain the Hamiltonian for the projected 2D brane $H_{\rm PB}$ from the 3D cubic lattice-based Hamiltonian, see Eq.~\eqref{eq:PTB}. The fourth column shows the number of sites on the one-dimensional (1D) brane $N^{\rm 1D}_s$ oriented along the body-diagonal of the parent cubic lattice of linear dimension $L$ in each direction. The fifth column shows the dimensionality of the matrix $H_{22}$, denoted by ${\rm dim}.(H^{3 \to 1}_{22})$, that needs to be inverted to obtain the effective Hamiltonian for the 1D brane from the parent 3D cubic lattice. Here, we do not account for two orbital degrees of freedom present in the Hamiltonian for the Weyl systems. 
}~\label{tab:matrixdimension}
\end{table}

Analogous to the parent disordered 3D Weyl system, the dirty 2D PB also exhibits a semimetal-to-metal quantum phase transition. This transition is primarily identified through the scaling of the ADOS at zero energy as a function of the disorder strength $W$: for weak disorder, the ADOS at zero energy remains pinned at zero, while beyond a critical disorder strength it becomes finite, as shown in Fig.~\ref{fig:PBWeylADOS}. Remarkably, the numerically obtained critical exponents for the semimetal-to-metal transition are nearly identical for both 3D cubic lattice Weyl fermions and the 2D projected Weyl brane (PWB), within the numerical accuracy, as shown in Fig.~\ref{fig:exponents}. These observations strongly indicate that the 2D PWB hosts the same semimetal-to-metal quantum phase transition as its parent cubic lattice system.

Finally, we demonstrate that, upon further increasing the disorder strength, the Weyl metal on the PB undergoes a second quantum phase transition at much stronger disorder, transitioning into an Anderson insulating phase. This transition is signaled by the vanishing of the TDOS at zero energy. Notably, the TDOS at zero energy is also critical across the semimetal-to-metal quantum phase transition, as it vanishes in both the semimetallic and Anderson insulating phases but is finite in the metallic phase. The critical exponent associated with the Anderson metal-insulator transition in the PWB is in reasonable agreement with that of the 3D cubic lattice Weyl model; see Fig.~\ref{fig:WeylAnderson}.

Thus, our results demonstrate that PBs inherit not only the topological features of their parent systems~\cite{panigrahi2022projected, tynerjuricic2024}, but also the complete disorder-driven quantum phase diagram and associated critical exponents. This finding defies the conventional expectation that robust metallic and semimetallic phases cannot exist in 2D non-interacting systems or can only be stable in the presence of strong interaction and weak disorder~\cite{ChakravartyAbrahams1998}. The quantitatively established dimensional inheritance reported here constitutes a significant conceptual advance and paves the way for realizing higher-dimensional quantum phenomena in experimentally accessible lower-dimensional platforms. With the main results summarized here and the necessary background established in the preceding subsections, we now proceed to present the technical details in the following sections.

\subsection{Organization}

The remainder of the manuscript is organized as follows. Section~\ref{sec:andersonmodel} is devoted to the discussion on the conventional Anderson model on 3D and 2D hypercubic lattices as well as on the 2D hexagonal PB. Stability of 3D and 2D Weyl fermions, and subsequently that of PWBs in two dimensions in the presence of randomness are studied in length in Sec.~\ref{sec:dirtyweyl}. We summarize the findings, highlight possible future directions, and present discussions on related topics, including possible experimental platforms to test our predictions in Sec.~\ref{sec:summary}. Additional technical details and results are relegated to two appendices. Namely, in Appendix~\ref{append:zeromodes} we detail the existence of exact zero-energy modes in our cubic-symmetric model for Weyl fermions due to quantization of momentum. Scaling and critical properties of disordered Anderson and Weyl models in $d=3$, obtained from kernel polynomial method (KPM) in large cubic lattices are shown in Appendix~\ref{append:KPM} to validate the exact diagonalization procedure in smaller systems, employed in the main manuscript.

\section{Anderson model}~\label{sec:andersonmodel}

The Anderson model has been the foundation for our understanding of the role of disorder on electronic fluids and localization~\cite{Anderson1958, thouless1970, AALR1979, Wegner1979, Lee1981, Hikami1981, RevModPhys1985AT, Kirkpatrick1994, janssen1998, slevin1999, slevin2001, klopp2002, dobrosavljevic2003, Brndiar2006, Chakravarty2008, RevModPhys2008AT, slevin2010}. In the site-localized or Wannier basis the 3D Anderson model (AM) can be written as 
\begin{equation}\label{eq:anderson}
    H_{\rm AM}=t\sum_{\langle i, j \rangle}|i\rangle \langle j| + \sum_{i}V(i)|i\rangle \langle i|,
\end{equation}
where $i=(x,y,z)$ is a site on a cubic lattice of linear size $L$ in each direction, $\langle i, j \rangle$ indicates a sum over the nearest neighbors site, $|i\rangle$ is the site-localized Wannier wave-function on the $i$th site, and we fix $t=1/2$. We model the quenched disorder as a random on-site potential $V(i)$. We sample the potential $V(i)$ from a Gaussian distribution with zero mean and standard deviation $W$, such that the probability density function is given by 
\begin{equation}~\label{eq:gaussiandis}
    f(x)=\frac{1}{\sqrt{2\pi W^2}} \: \exp\left( -\frac{x}{2W^2} \right).
\end{equation}
 As a result, $W$ characterizes the strength of disorder. In order to establish the phase diagram in the presence of disorder we compute the TDOS that at any given energy $E$ is defined as 
\begin{equation}~\label{eq:tdos}
\rho_{\mathrm{typ}}(E)=\text{exp}\left(\frac{1}{N_t}\sum_{i=1}^{N_t} \langle \langle \ln\rho_{i}(E) \rangle \rangle \right),
\end{equation} 
where $N_t \ll L^d$ indicates a small collection of random sites considered for each disorder configuration, $\langle \langle \cdots \rangle \rangle$ represents the disorder averaged quantity, and the local DOS at site $i$ and at energy $E$ is defined as 
\begin{equation}~\label{eq:LDOS}
    \rho_{i}(E)=\sum_{j}|\langle j|i \rangle|^2\delta(E-E_{j}).
\end{equation}
The TDOS serves as an order parameter for localization, because it represents the geometric average of the local DOS. When the electronic states localize, the local spectrum changes from being continuous to discrete. This has the effect of suppressing the geometric average and lowering the TDOS, making it a suitable quantity for tracking the localization of electronic wave functions across the Anderson transition. All the corresponding numerical analyses are performed using exact diagonalization and we always impose periodic boundary conditions in all directions of the $d$-dimensional hypercubic lattice, which gets inherited by the PBs.

We compute the TDOS at zero energy from Eq.~\eqref{eq:anderson} after averaging over 500 independent disorder configurations as a function of the disorder strength on a cubic system of linear size $L$ with the results shown in Fig.~\ref{fig:BraneAnderson}(a). We note that the metallic phase, characterized by a finite $\rho_{\rm typ}(0)$, is stable for weak disorder, showing minimal deviation as the system size is increased. However, at a large disorder the TDOS at $E=0$ vanishes as expected, indicating a metal-insulator transition, beyond which the system describes an Anderson insulator. In addition, by fitting the TDOS at zero-energy with the following single-parameter scaling ansatz 
\begin{equation}~\label{eq:Andersonscaling}
\rho_{\rm typ}(0) \sim \left( \frac{W_c-W}{W_c}\right)^{\beta},
\end{equation}
where $W_c=3.50 \pm 0.15$ is the critical disorder strength for the Anderson transition, we numerically compute the corresponding order parameter exponent $\beta$ in a $L=20$ cubic lattice, yielding $\beta=1.60 \pm 0.14$; see Fig.~\ref{fig:BraneAnderson}(b). The value of this exponent is reasonably close to that reported in the literature ($\beta \sim 1.55$)~\cite{slevin2001}. To benchmark these outcomes, we also compute the TDOS at zero energy for the Anderson model in much larger cubic lattices using KPM, detailed in Appendix~\ref{append:KPM}, yielding $W_c=3.10 \pm 0.10$ and $\beta=1.60 \pm 0.16$ on a $L=60$ cubic lattice.

In two spatial dimensions, the Anderson model is identical to Eq.~\eqref{eq:anderson}, however with $i=(x,y)$. Using this two-dimensional model, we recompute the TDOS at zero energy for two sufficiently weak disorder strengths $W=0.20$ and $0.40$, as a function of the system size $L$ (linear dimension of the square lattice with periodic boundary condition in each direction) after averaging over 500 independent disorder configurations. We then plot the TDOS at zero energy as a function of $1/L$ and perform a linear fit of the data to extrapolate to the thermodynamic limit $L \to \infty$. We note that TDOS at zero energy tends to zero prior to reaching the true thermodynamic limit, as shown in Fig.~\ref{fig:BraneAnderson}(c), indicating an absence of a stable metallic phase in two dimensions. We also note that the linear fit of the TDOS at zero-energy with $1/L$ approaches zero for smaller $1/L$ or equivalent larger $L$ with decreasing disorder strength, confirming that stronger disorder induces localization in smaller systems.

\subsection{Anderson model on 2D projected brane}

Having established the distinct responses of three- and two-dimensional metallic phases to disorder within the Anderson model, we next set out to establish stability and quantum criticality of a 2D dirty PB constructed from the 3D Anderson model following the procedure described in Sec.~\ref{subsec:constructionPB} with $H_{\rm parent}=H_{\rm AM}$ in Eq.~\eqref{eq:parent}. Upon constructing $H_{\rm PB}$ for the 2D PB from a cubic lattice of linear dimension $L$, we introduce the onsite potential $V(i)$ therein, drawn from a random Gaussian distribution with zero mean and standard deviation $W$, see Eq.~\eqref{eq:gaussiandis}.

\begin{figure}
\includegraphics[width=1.00\linewidth]{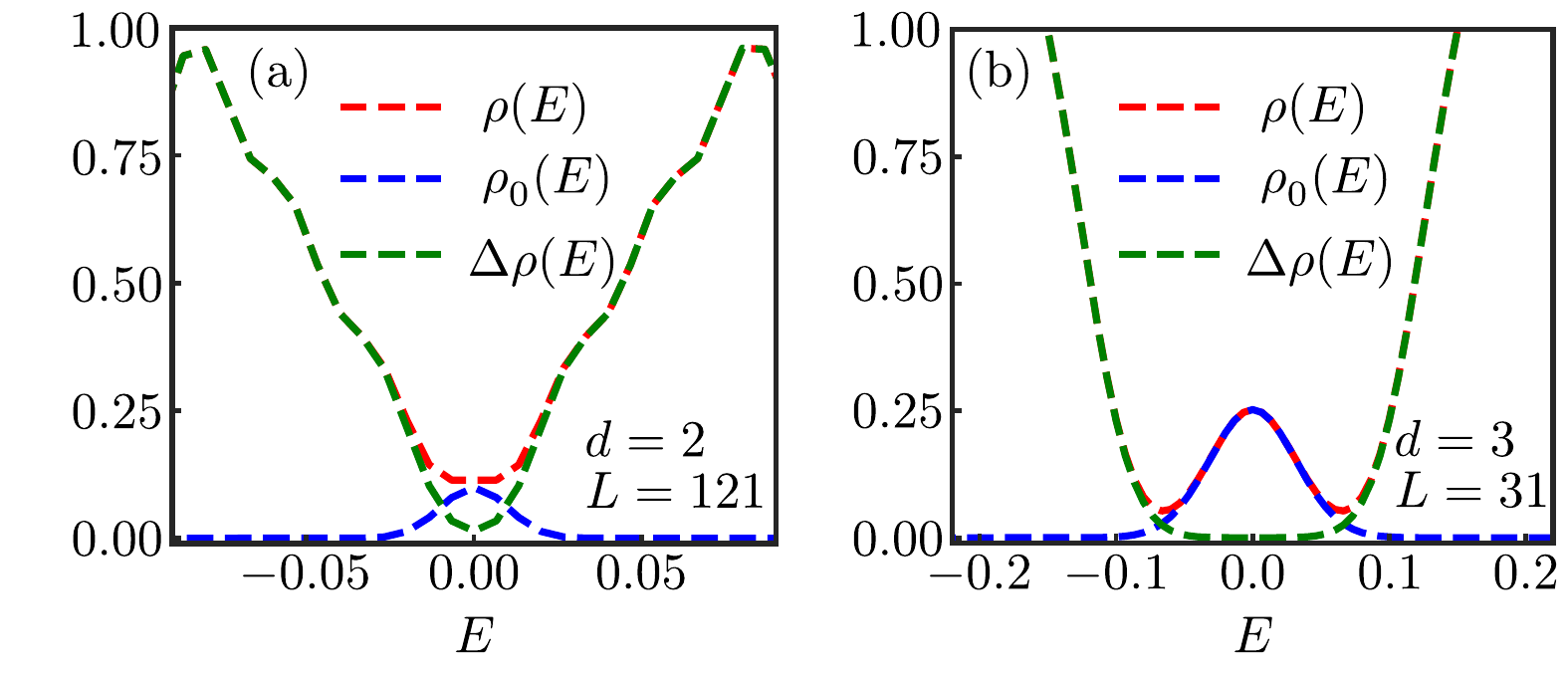}
     \caption{Total density of states (DOS) $\rho(E)$ (red), DOS solely due to the zero-energy modes $\rho_0(E)$ (blue), and the subtracted DOS $\Delta \rho(E)=\rho(E)-\rho_0(E)$ (dark green) for (a) a two-dimensional ($d=2$) Weyl Hamiltonian on a square lattice of linear dimension $L=131$ and (b) a three-dimensional ($d=3$) Weyl Hamiltonian on a cubic lattice of linear dimension $L=31$ and periodic boundary condition in each direction in clean systems. For definitions of these quantities, see Eqs.~\eqref{eq:subtractedDOS}-\eqref{eq:ados3}. The hypercubic symmetric Weyl Hamiltonian in $d$ dimension is introduced in Eq.~\eqref{eqn:fullH}.  
     }~\label{fig:DOSsubtraction}
\end{figure}

The TDOS at zero energy on PB is computed following Eq.~\eqref{eq:tdos} after averaging over 500 independent disorder configurations. At this point, an important comment is due. The familiar sparse matrix methods such as the KPM~\cite{kpm} and Lanczos~\cite{lanczos1950}, which allow access to sufficiently large system sizes, cannot be implemented  for PBs due to the need to compute the inverse of $H_{22}$, see Eq.~\eqref{eq:PTB}. While each entry of the parent Hamiltonian $H_{\rm parent}$, namely $H_{11}$, $H_{22}$, $H_{12}$, and $H_{21}$, is indeed always a sparse matrix in all our constructions with hopping amplitudes only among the nearest-neighbor sites and on-site potential disorder, the inverse of a spare matrix is generally non-sparse. As a result, the inverse of $H_{22}$ and therefore $H_{\rm PB}$ is \emph{not} a sparse matrix. The numerical calculation of the matrix inverse is a computationally demanding procedure and has proven to be a limiting factor in our consideration of increasingly larger system sizes of the PB, characterized by the total number of sites therein $N^{\rm 2D}_s$. As a result, we always rely on exact diagonalization. Table~\ref{tab:matrixdimension} illustrates how the dimensionality of $H_{22}$, which must be inverted to construct \(H_{\rm PB}\), increases rapidly with $L$ and the PB system size $N^{\rm 2D}_s$.

The results for the Anderson model on 2D hexagonal PB, shown in Fig.~\ref{fig:BraneAnderson}(d), demonstrate a stable metallic phase over a large range of disorder, which remains unaffected as $N^{2D}_s$ is increased. As disorder strength is increased further, we observe a metal-insulator transition at disorder strength $W_c=17.95 \pm 0.50$, in alignment with the behavior of the 3D cubic lattice Anderson model, rather than its 2D counterpart on a square lattice. Furthermore, from the scaling ansatz of the TDOS at zero-energy near the Anderson transition from Eq.~\eqref{eq:Andersonscaling}, we obtain the corresponding order parameter exponent to be $\beta=1.55 \pm 0.10$, reasonably close to the one we reported for the 3D cubic lattice Anderson model. See Fig.~\ref{fig:BraneAnderson}(e). These findings strongly suggest that the 2D PB, constructed via the Schur complement, can capture the quantum phase diagram and associated critical properties of the parent 3D cubic lattice. To further anchor this claim, next we proceed to analyze the global phase diagram of a PWB, constructed from its parent Weyl Hamiltonian on a cubic lattice in the presence of random on-site charge impurities.

\section{Dirty Weyl fermions}~\label{sec:dirtyweyl}

Dirac and Weyl semimetals are characterized by gapless excitations at the Fermi energy, but only around a few isolated points in the Brillouin zone (thus also named nodal Fermi liquids). These semimetals are of intense theoretical and experimental interest as they can be found as a stable phase of matter protected by crystalline symmetries as well as at the critical point separating two topologically distinct insulating phases~\cite{Herring1937, Jia2016, ChiuRMP2016, Yan2017, ArmitageRMP2018}. In addition, Weyl fermions can be found in weakly interacting materials, inside magnetically ordered phases in strongly correlated materials~\cite{Savrasov2011, YingRan2011, YamajiImada2014, WitczakKrempa2014, Savary2014, GoswamiRoyDasSarma2017, LadovrechisMengRoy2021}, and in Kondo materials~\cite{Lai2017, Chen2022, Dzsaber2017}. Due to their correspondence with topology and topological quantum critical points, the stability criteria of Weyl and Dirac semimetals against disorder~\cite{Fradkin1986, Murakami2009, goswami2011, Imura2013, Herbut2014, roydassarma2014, Syzranov2015a, Syzranov2015b, Altland2015, pix2015, Syzranov2016, Pix2016a, roydassarma2016, bera2016, Gorbar2016, RoyJuricicDasSarma2016, RoyDasSarma2016PRB, Carpentier2016, Carpentier2017, Justin2017, Goswami2017a, Goswami2017b, Alavirad2017, Slager2017, WilsonRefael2018, Slager2018, Carpentier2018, Carpentier2019, Szabo2020, Gruzberg2024, Huse2014, Huse2016a, Huse2016b, buch2018, JustinWilson2020, JustinWilson2024, PixleyWilsonReview2021} and interaction~\cite{Hosur2012PRL, VivekAji2012, IsobeNagaosa2012, WangZhang2013, MaciejkoNandkishore2014, HsinHuaLai2015, HongYao2015, HofmannPRB2015, WangPeng2016, Laubach2016, Katsnelson2016, roy2017, RoyJuricic2017PRB, Gonzalez2017, BergholtzPRB2018, Malve2025} have been studied extensively.

We first consider the case of familiar 3D and 2D Weyl semimetals. On a $d$-dimensional hypercubic lattice of lattice spacing $a$ in each direction, the Weyl Hamiltonian takes the following generic form
\begin{equation}~\label{eqn:fullH}
     H=\frac{t}{2i} \sum_{\vec{r}} \sum_{j=1,\cdots,d} \Psi^\dagger_{\vec{r}} \tau_j \Psi_{\vec{r}+\hat{e}_j}
     + \sum_{\vec{r}} V (\vec{r}) \Psi^\dagger_{\vec{r}} \Psi_{\vec{r}},  
\end{equation}
where $\hat{e}_j= a \hat{j}$ with $\hat{j}$ as the unit vector along $j=1,\cdots,d$. The mutually anticommuting matrices, constituting the set $\{ \tau_j\}$, operate on the orbital indices with opposite parities. Therefore, in $d=2$ and $3$, they correspond to Pauli matrices. Then, $\Psi^\top_{\vec{r}}=(c_{\vec{r},+}, c_{\vec{r},-})$ is a two-component spinor, where $c_{\vec{r},\tau}$ is the fermionic annihilation operator at position $\vec{r}=(x,y,z)$ in $d=3$ and $\vec{r}=(x,y)$ in $d=2$ with parity eigenvalue $\tau=\pm$. In $d=3$ ($d=2$), the above tight-binding model in clean system supports linearly dispersing Weyl fermions around eight (four) high-symmetry and time-reversal invariant momentum points of the corresponding Brillouin zone.

\begin{figure}
\includegraphics[width=1.00\linewidth]{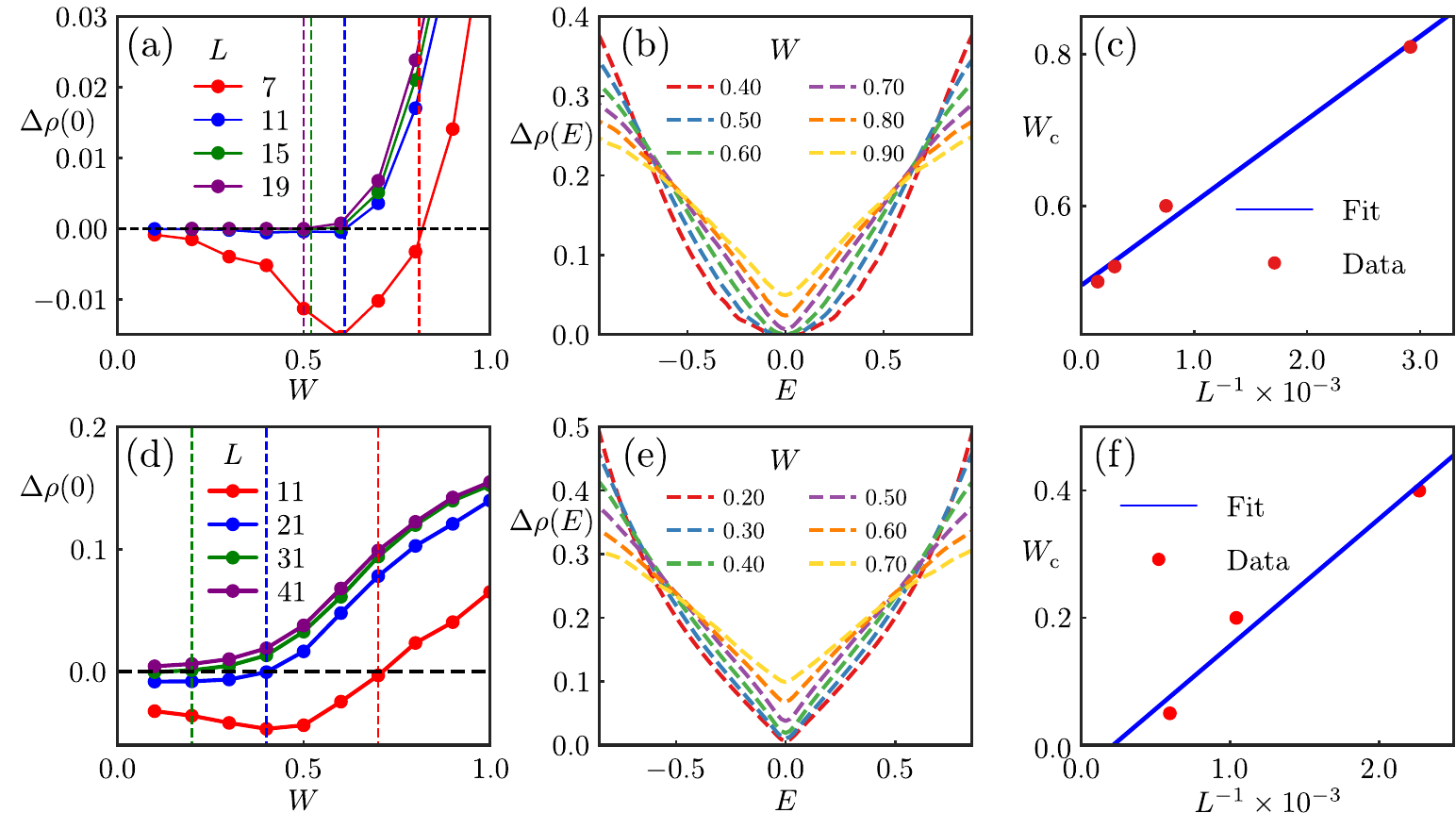}
     \caption{(a) Dependence of the subtracted average density of states (ADOS) at zero-energy $\Delta \rho(0)$ [see Eq.~\eqref{eq:subtractedDOS}] with the disorder strength ($W$) for a three-dimensional (3D) Weyl semimetal as a function of the linear dimension of the cubic lattice $L$ in each direction (see legend). (b) Subtracted ADOS $\Delta \rho(E)$ as a function of $W$ for a 3D Weyl system on a cubic lattice with $L=19$. (c) Critical value  of disorder ($W_{c}$) for the semimetal-to-metal quantum phase transition as a function of $L^{-1}$, demonstrating the convergence of $W_c$ to a finite value of $0.49 \pm 0.05$ in the thermodynamic limit ($L^{-1} \to 0$). (d) Variation of $\Delta \rho(0)$ with $W$ for a two-dimensional (2D) Weyl semimetal as a function of the linear dimension of the square lattice $L$ in each direction. (e) Scaling of $\Delta \rho(E)$ for a 2D disordered Weyl semimetal as a function of $W$ for a system with $L=41$. (f) Critical value  of disorder $W_{c}$ for a \emph{putative} semimetal-to-metal transition for a 2D Weyl system as a function of $L^{-1}$, showing that $W_c \to 0$ even before the system size reaches the true thermodynamic limit, confirming the absence of such a transition in $d=2$. Here, all data points are obtained from exact diagonalization after averaging over 500 independent disorder realizations. We always impose periodic boundary conditions in all directions.   
     }~\label{fig:DOSWeyl2D3D}
\end{figure}

We again model the quenched disorder as a random on-site potential $V(\vec{r})$, sampled from a Gaussian distribution with zero mean and standard deviation $W$, see Eq.~\eqref{eq:gaussiandis}. In order to quantify the effects of disorder and distinguish the semimetallic and metallic phases we compute the ADOS at energy $E$, defined as 
\begin{equation}\label{eq:ADOS}
\rho_{\rm avg}(E)=\frac{1}{N} \left[ \sum_{n}\delta(E-E_n) \right],
\end{equation}
where the summation is over all the eigenstates $|E_n\rangle$ with energy $E_n$ of the disordered Hamiltonian and $N$ is the total number of states in the system. For the $d$-dimensional Hamiltonian on a hypercubic lattice, $N=2L^{d}$, where the factor of $2$ arises from the orbital degrees of freedom. In the clean limit, this tight-binding model is known to support two zero energy states under periodic boundary conditions for linear dimension of the system $L=2M+1$, where $M \in \mathbb{Z}$ (integer). The presence of such zero energy states can, in principle, obscure the semimetal-to-metal transition as a function of the increasing disorder strength. The origin of these zero modes is detailed in Appendix~\ref{append:zeromodes}. To overcome this challenge prior works have utilized a number of methods including twisted boundary conditions to create a small gap in the spectra~\cite{Huse2016a}. In this work, we employ a rather straightforward procedure in which, when extracting the ADOS, we subtract the contribution due to such zero-energy modes, computed in the clean system $\rho^{\rm clean}_{\rm avg}(0) \equiv \rho_{\rm avg}(0)$. We therefore track the \emph{subtracted} ADOS, defined as 
\begin{equation}~\label{eq:subtractedDOS}
\Delta \rho(E)=\rho_{\rm avg}(E)-\rho_{\rm avg}(0), 
\end{equation}
as a function of disorder strength ($W$) to identify the semimetal-metal quantum phase transition, where  
\begin{equation}~\label{eq:ados2}
    \rho_{\rm avg}(E)= \frac{1}{2N\sqrt{\epsilon\pi}} \: \left[ \; \sum_{n} \; \exp\left( -\frac{(E-E_{n})^2}{4\epsilon} \right) \right],
\end{equation}
with $\epsilon$ as the broadening parameter, which we set to be $5\times 10^{-4}$ throughout, and 
\begin{equation}~\label{eq:ados3}
    \rho_{\rm avg}(0)= \frac{1}{N \sqrt{\epsilon \pi}} \: \exp\left(-\frac{E^2}{4\epsilon} \right).
\end{equation}
At this point it is clear that the effect of the zero-energy modes is eliminated in the thermodynamic limit as the number of zero modes is fixed for even or odd $L$, while the contribution of the zero modes to the DOS scales down as $1/N$. To illustrate this method, we show $\Delta \rho(E)$, $\rho_{\rm avg}(E)$, and $\rho_{\rm avg}(0)$ in Fig.~\ref{fig:DOSsubtraction}(a) for a 2D clean Weyl semimetal of $L=121$ and Fig.~\ref{fig:DOSsubtraction}(b) for a 3D clean Weyl semimetal of $L=31$. While this procedure ensures that $\Delta \rho(0)=0$ in clean Weyl systems, their characteristic $|E|$ and $|E|^2$ scaling of the DOS in $d=2$ and $3$, respectively, can only be observed slightly away from the zero-energy. However, with the introduction of small disorder as these modes move slightly away from zero-energy, we recover the desired power-law scaling of the ADOS. In the presence of disorder, we compute $\rho_{\rm avg}(E)$ after disorder averaging, while $\rho_{\rm avg}(0)$ is \emph{always} computed in a clean system and we investigate the scaling of disorder-averaged subtracted ADOS $\Delta \rho(E)$ from Eq.~\eqref{eq:subtractedDOS} as a function of the disorder strength $W$ to identify the semimetal-to-metal quantum phase transition, which we discuss next.

In general, weak disorder causes a minor shift in energy to these otherwise zero-energy modes present in the clean systems. As a result, it is expected that, while in the clean limit $\Delta \rho(0)=0$, for weak disorder such that the system remains in a semimetallic phase with $\Delta \rho(0)<0$. Then we can identify the semimetal-to-metal critical point as the value of disorder strength $W_{c}$ at which $\Delta \rho(0)$ becomes \emph{positive}. This transition is visible in Fig.~\ref{fig:DOSWeyl2D3D}(a) where $\Delta \rho(0)$ is plotted as a function of the disorder strength for multiple values of the linear system size $L$ of the cubic lattice. We further demonstrate $\Delta \rho(E)$ as a function of disorder strength in Fig.~\ref{fig:DOSWeyl2D3D}(b) for the largest cubic lattice considered in this work for the exact diagonalization with $L=19$. We note that as a function of increasing system size, the value of $W_{c}$ converges to a fixed value of $0.49 \pm 0.05$ indicating the stability of the semimetallic phase to the introduction of weak potential disorder and its transition to a metallic phase at moderate disorder ($W_c$), as shown in Fig.~\ref{fig:DOSWeyl2D3D}(c). These findings are consistent with the extensive previous analytical and numerical works that have been performed over the years~\cite{Fradkin1986, Murakami2009, goswami2011, Imura2013, Herbut2014, roydassarma2014, Syzranov2015a, Syzranov2015b, Altland2015, pix2015, Syzranov2016, Pix2016a, roydassarma2016, bera2016, Gorbar2016, RoyJuricicDasSarma2016, RoyDasSarma2016PRB, Carpentier2016, Carpentier2017, Justin2017, Goswami2017a, Goswami2017b, Alavirad2017, Slager2017, WilsonRefael2018, Slager2018, Carpentier2018, Carpentier2019, Szabo2020, Gruzberg2024}.

The scaling analysis shown in Fig.~\ref{fig:DOSWeyl2D3D}(c) as a function of system size is crucial due to the previously stated computational limitations in the construction of the PB. However, to benchmark the scaling analysis we utilize the KPM to compute ADOS for the Hamiltonian from Eq.~\eqref{eqn:fullH} on a cubic lattice of a large linear dimension of $L=121$. Full details of the calculation are provided in Appendix~\ref{append:KPM}. We note that such a calculation yields $W_{c}=0.50 \pm 0.05$ for the Weyl semimetal-to-metal quantum phase transition, in excellent agreement with the value of $W_c$ obtained from the scaling of the subtracted ADOS analysis, yielding $W_{c}=0.49 \pm 0.05$.

\begin{figure}
\includegraphics[width=1.00\linewidth]{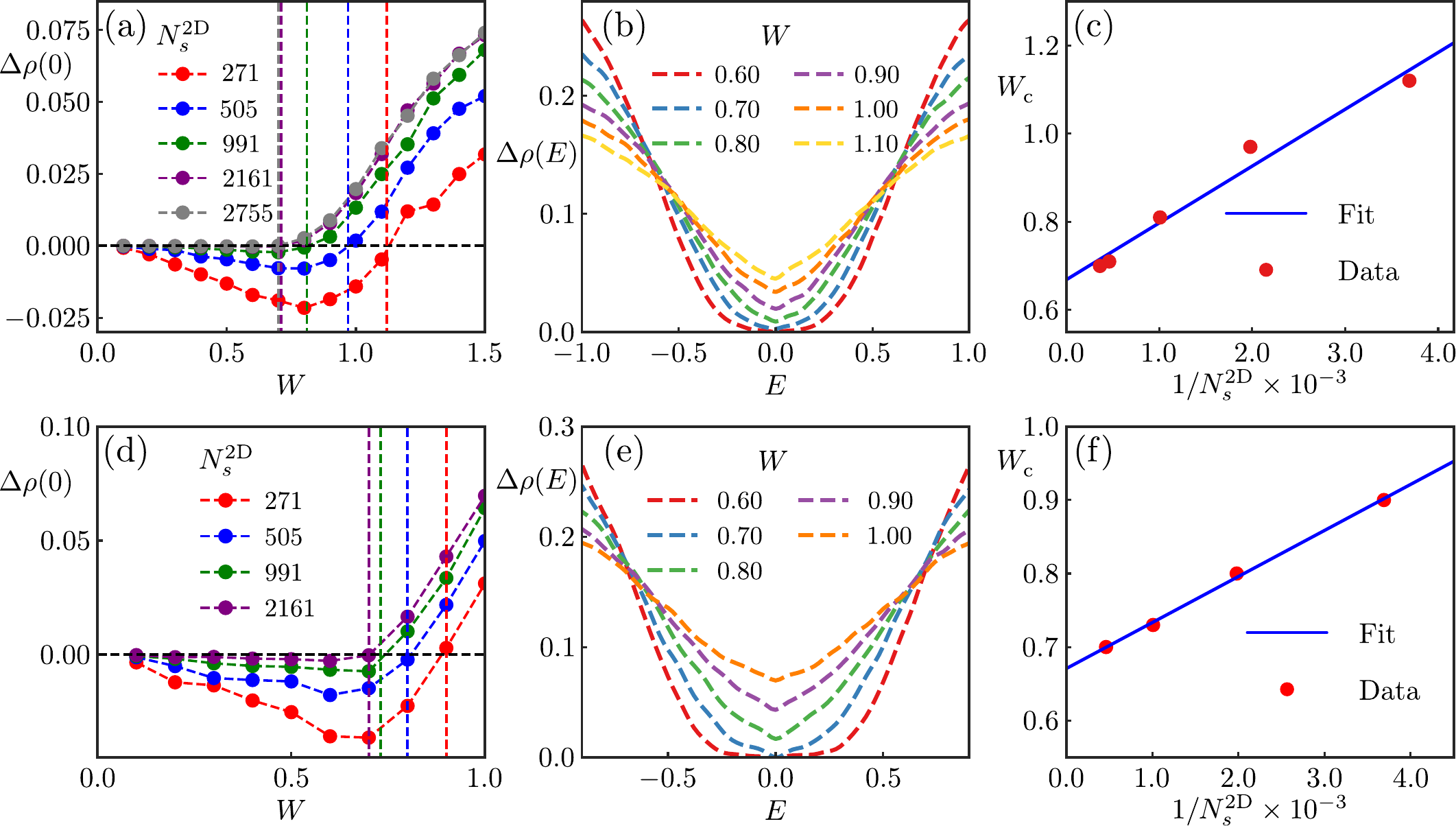}
     \caption{(a) Dependence of the subtracted average density of states (ADOS) at zero-energy $\Delta \rho(0)$  [see Eq.~\eqref{eq:subtractedDOS}] with the disorder strength ($W$) for a two-dimensional (2D) projected Weyl brane (PWB) as a function of the number of sites therein $N^{\rm 2D}_s$ (see legend). (b) Variation of $\Delta \rho(E)$ as a function of $W$ for a 2D PWB with $N^{\rm 2D}_s=2755$. (c) Critical value  of disorder ($W_{c}$) for the semimetal-to-metal quantum phase transition as a function of $(N^{\rm 2D}_s)^{-1}$, demonstrating the convergence of $W_c$ to a finite value of $0.70 \pm 0.02$ in the thermodynamic limit $(N^{\rm 2D}_s)^{-1} \to 0$. These results are obtained after projecting the clean cubic lattice Weyl Hamiltonian onto the brane and subsequently introducing disorder therein. Panels (d)-(f) are analogous to (a)-(c), respectively, but the results are obtained by projecting disordered cubic lattice Weyl Hamiltonian onto the brane. As shown in (f), the estimated critical disorder in this case is also $W_c =0.70 \pm 0.02$ as $(N^{\rm 2D}_s)^{-1} \to 0$. Here, all data points are obtained from the exact diagonalization after averaging over 500 independent disorder realizations and we always impose periodic boundary conditions in all directions of the parent cubic lattice, which is inherited by PWBs.   
     }~\label{fig:PBWeylADOS}
\end{figure}

Next, we turn the focus onto the role of disorder on 2D Weyl fermions realized on a square lattice. The corresponding Hamiltonian can readily be obtained from Eq.~\eqref{eqn:fullH} upon setting $d=2$ therein. The remainder of the analysis follows the same procedure discussed above for Weyl fermions on a cubic lattice, and the corresponding results are shown in Fig.~\ref{fig:DOSWeyl2D3D}(d)-\ref{fig:DOSWeyl2D3D}(f). These findings are consistent with previous conclusions indicating that the semimetallic phase in $d=2$ is unstable to infinitesimal disorder, as evidenced by $\Delta \rho(0) > 0$ for any finite disorder strength $W$, provided a sufficiently large system size is considered. Having recovered the known results for two- and three-dimensional Weyl fermions on hypercubic lattices using our approach, we now proceed to investigate the effects of disorder on the 2D PWBs following the same procedure: computation of disorder-averaged subtracted ADOS from exact diagonalization.

\subsection{Dirty 2D projected Weyl brane}

Having established a marked contrast in the stability of two- and three-dimensional Weyl semimetals under weak potential disorder in the previous section, we now turn to the effects of disorder on a 2D PWB. This PWB is constructed from Eq.~\eqref{eqn:fullH}, following the general principle outlined in Eq.~\eqref{eq:PTB}, for a parent cubic lattice of linear size $L$, as detailed in Sec.~\ref{subsec:constructionPB}. The results for $\Delta \rho(0)$, obtained by averaging over 500 independent disorder configurations, as a function of system size ($N^{\rm 2D}_s$) and disorder strength are presented in Fig.~\ref{fig:PBWeylADOS}(a). Correspondingly, $\Delta \rho(E)$ as a function of disorder strength is shown in Fig.~\ref{fig:PBWeylADOS}(b). These data demonstrate that the critical disorder strength converges to $W_{c} = 0.70 \pm 0.02$ with increasing $N^{\rm 2D}_s$ as the thermodynamic limit is approached ($1/N^{\rm 2D}_s \to 0$), as illustrated in Fig.~\ref{fig:PBWeylADOS}(c), thereby establishing the robustness of the semimetallic phase despite the reduced physical dimensionality of the system. These results are obtained by first projecting the clean Weyl Hamiltonian onto the 2D brane and subsequently introducing disorder therein.

\begin{figure}
\includegraphics[width=1.00\linewidth]{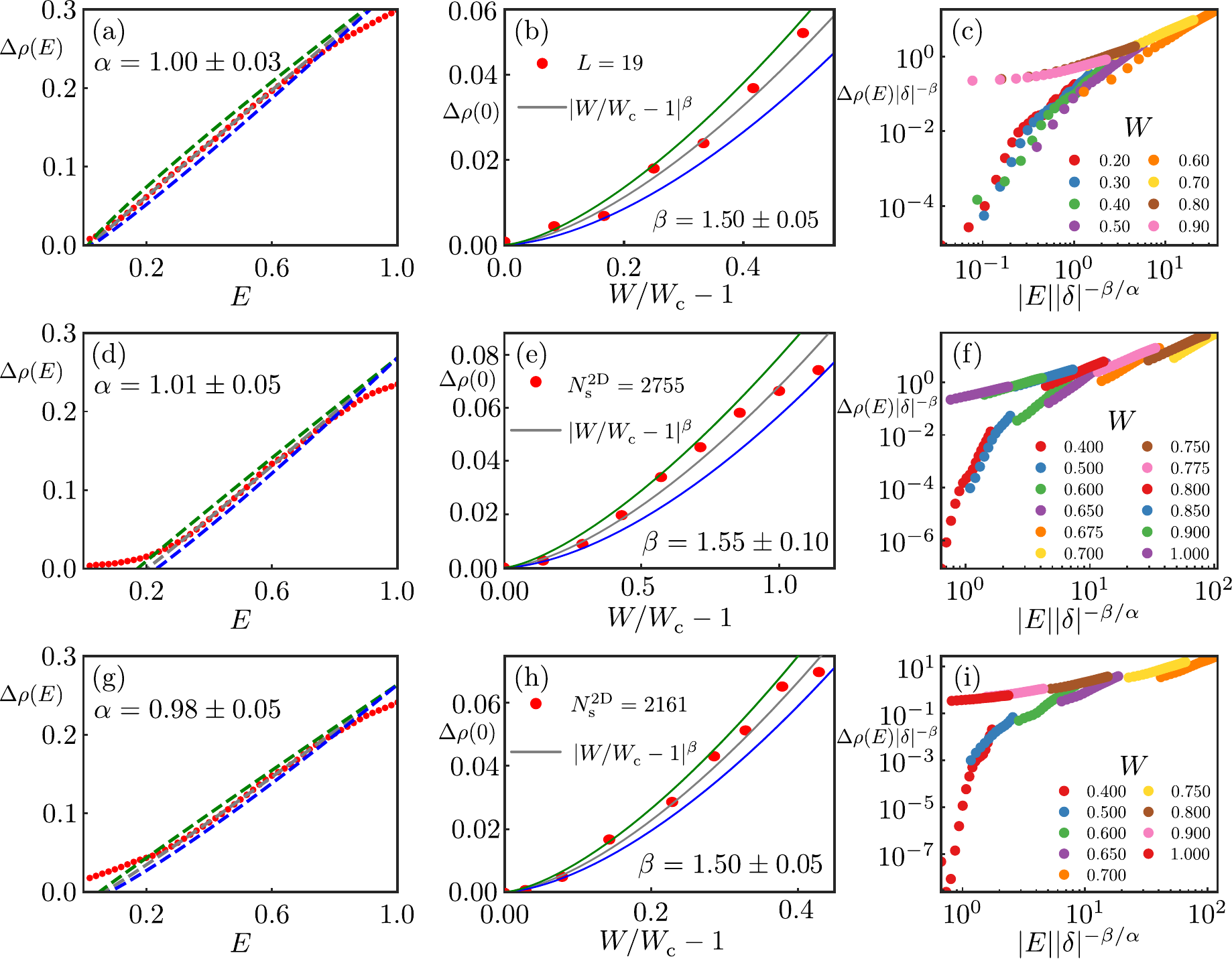}
     \caption{Computation of (a) average density of states exponent $\alpha$, obtained from the scaling function $\rho(E) \sim |E|^{\alpha}$ near the critical disorder strength  for the semimetal-to-metal transition and (b) order parameter exponent $\beta$, obtained from the scaling function $\rho(0) \sim |\delta|^{\beta}$ in the metallic phase of the same transition for a three-dimensional cubic lattice Weyl Hamiltonian with $L=31$ (linear dimension of the system in each direction). (c) Data collapse in the same system obtained by comparing $|E| \delta^{-\beta/\alpha}$ and $\Delta \rho(E) \delta^{-\beta}$, with the numerically obtained values of $\alpha=1.00$ and $\beta=1.55$. Subfigures (d)-(f) are analogous to (a)-(c), respectively, but for the projected Weyl brane (PWB) with $N^{\rm 2D}_s=2755$ (number of sites therein), obtained from a parent cubic lattice Weyl Hamiltonian with $L=35$. These results are obtained by introducing disorder to the otherwise clean system Hamiltonian for the PWB and in (f) data collapse are shown for $\alpha=1.01$ and $\beta=1.55$. Subfigures (g)-(i) are analogous to (d)-(f), but for the PWB with $N^{\rm 2D}_s=2161$ for which the effective Hamiltonian is constructed by projecting disordered Weyl Hamiltonian on a cubic lattice with $L=31$. In (i), data collapse is shown for $\alpha=0.98$ and $\beta=1.50$. See Sec.~\ref{subsec:Weylscaling} for details.   
     }~\label{fig:exponents}
\end{figure}

Analogous analyses are carried out by directly projecting the disordered Weyl Hamiltonian onto the 2D brane with the corresponding results shown in Figs.~\ref{fig:PBWeylADOS}(d)–\ref{fig:PBWeylADOS}(f). This approach once again yields $W_c = 0.70 \pm 0.02$ as the thermodynamic limit is approached, consistent with the previous construction. This comperative analysis is important for the following reason. Even when we introduce on-site disorder in 3D cubic lattice, under the projection procedure they become long-ranged or correlated disorder on PWBs. Although correlated disorder often changes the location and universality class of disorder-driven transitions, that turns out not to be the case (within numerical accuracy) in our procedure of constructing the PWBs, which we further substantiate in the next subsection. Altogether, these findings unambiguously establish the stability of 2D PWBs against sufficiently weak disorder and their transition to a diffusive metallic phase at a finite disorder strength $W_c$, in direct analogy with the behavior observed in the parent cubic lattice-based Weyl systems in three dimensions. In the following, we proceed to investigate the quantum critical properties associated with the disorder-induced semimetal-to-metal transition from the scaling of the subtracted ADOS with disorder strength.

\subsection{Scaling near Weyl semimetal-to-metal transition}~\label{subsec:Weylscaling}

The scaling theory for the ADOS across the Weyl semimetal-to-metal quantum phase transition in $d=3$ is well established~\cite{Herbut2014, pix2015, bera2016, Slager2018}. We briefly review it and subsequently apply it to 2D PWBs. Defining the dimensionless reduced distance from the quantum critical point as $\delta = (W - W_c)/W_c$, the universal scaling form of the subtracted ADOS in $d$ spatial dimensions is given by
\begin{equation}\label{eq:scalingADOS}
\Delta \rho (E) = \delta^{\nu(d-z)} F\left( |E| \delta^{-\nu z} \right),
\end{equation}
where $F$ is an \emph{unknown} universal scaling function of its argument, and $z$ and $\nu$ are the dynamic scaling exponent and correlation length exponent, respectively. Here, we neglect the system size dependence of the scaling function. The independence of $\Delta \rho (E)$ with respect to $\delta$ at the critical point ($\delta=0$) yields the scaling form $\Delta \rho(E) \sim |E|^{d/z-1}$, from which we extract the DOS scaling exponent $\alpha = d/z - 1$. In the metallic phase, the finite value of $\Delta \rho(0)$ near the critical point follows the scaling $\Delta \rho(0) \sim \delta^{\nu(d-z)}$, allowing us to define the order-parameter exponent $\beta = \nu(d-z)$. While the values of $z$ and $\nu$ can be directly extracted from numerically computed $\alpha$ and $\beta$ for a cubic lattice-based disordered Weyl Hamiltonian with $d=3$, the effective dimensionality of the 2D PWB is, however, ambiguous. Therefore, we focus on reporting the two independent critical exponents $\alpha$ and $\beta$. Furthermore, the universal scaling function in Eq.~\eqref{eq:scalingADOS} can be recast in terms of these exponents as
\begin{equation}\label{eq:scalingADOSalt}
\Delta \rho(E) = \delta^{\beta}F(|E| \delta^{-\beta/\alpha}),
\end{equation}
which we use to demonstrate data collapse and single-parameter scaling, hallmarks of continuous or second-order phase transitions. The resulting analysis is summarized in Fig.~\ref{fig:exponents}, which we discuss next.

We first focus on the 3D Weyl system on a cubic lattice, obtaining $\alpha = 1.00 \pm 0.03$ [Fig.~\ref{fig:exponents}(a)] and $\beta = 1.50 \pm 0.05$ [Fig.~\ref{fig:exponents}(b)] from the aforementioned scaling forms. These results, derived via exact diagonalization on a small cubic lattice of $L=19$, are in excellent agreement with previously reported values from KPM in larger systems, which we also report in Appendix~\ref{append:KPM} on a $L=121$ cubic lattice, yielding $\alpha=1.00 \pm 0.05$ and $\beta=1.55 \pm 0.10$. This consistency justifies the use of exact diagonalization on smaller systems for extracting critical exponents near the Weyl semimetal-to-metal quantum phase transition (within numerical accuracy), which we next apply to the disordered 2D PWBs. Notably, comparing $|E| \delta^{-\beta/\alpha}$ against $\Delta \rho(E) \delta^{-\beta}$, following  Eq.~\eqref{eq:scalingADOSalt}, we observe that all data points collapse onto three distinct branches for $\alpha=1.00$ and $\beta=1.55$, as shown in Fig.~\ref{fig:exponents}(c), corresponding to the semimetallic phase (lower-left branch), the metallic phase (upper-left branch), and the quantum critical regime (upper-right branch), where the former two branches meet.

When the 2D PWB of $N^{\rm 2D}_s=2755$ is constructed from a clean 3D cubic lattice Weyl Hamiltonian on which we introduce random charge impurities, we find $\alpha = 1.01 \pm 0.05$ [Fig.~\ref{fig:exponents}(d)] and $\beta = 1.55 \pm 0.10$ [Fig.~\ref{fig:exponents}(e)]. Here, all the data collapse onto three branches with $\alpha = 1.01$ and $\beta = 1.55$, as shown in Fig.~\ref{fig:exponents}(f). By contrast, for disordered 2D PWBs of $N^{\rm 2D}_s=2161$ constructed by projecting the disordered 3D Weyl Hamiltonian, we obtain $\alpha = 0.98 \pm 0.05$ [Fig.~\ref{fig:exponents}(g)] and $\beta = 1.50 \pm 0.10$ [Fig.~\ref{fig:exponents}(h)], again yielding a collapse of all data points onto three branches for $\alpha=0.98$ and $\beta=1.50$ [Fig.~\ref{fig:exponents}(i)].

These findings demonstrate that 2D dirty PWBs not only share the same quantum phase diagrams as disordered 3D semimetals, but also exhibit \emph{identical} quantum critical behavior across the semimetal-to-metal transition (within numerical accuracy). Furthermore, this conclusion does not depend on whether disorder is introduced on Weyl branes at after projecting the clean 3D Weyl Hamiltonian yielding local or on-site disorder therein or we project disordered 3D Weyl Hamiltonian, yielding long-range or correlated disorder on PWBs. Thus, 2D PWBs can be regarded as genuine quantum holographic images of their parent 3D Weyl materials. In the following section, we further substantiate this claim by analyzing the scaling of the TDOS in both systems, which, besides the semimetal-to-metal transition, also reveals the Anderson metal-insulator transition in Weyl systems.

\begin{figure}
\includegraphics[width=1.00\linewidth]{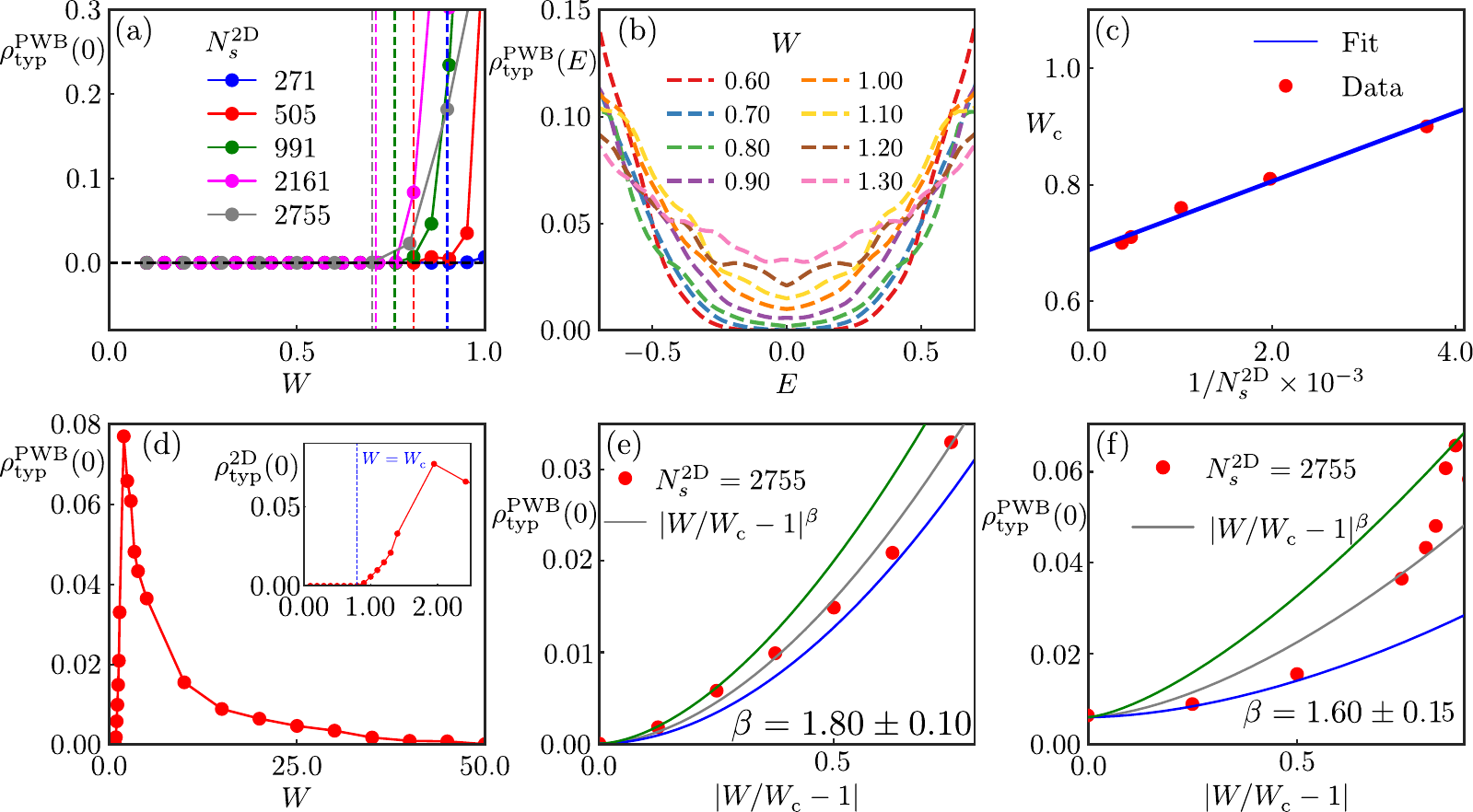}
     \caption{(a) Dependence of the typical density of states (TDOS) at zero energy $\rho^{\rm PWB}_{\rm typ} (0)$ on the disordered two-dimensional (2D) projected Weyl brane (PWB) with disorder strength ($W$) as a function of the number of lattice sites ($N^{\rm 2D}_s$) therein (see legend). See Table~\ref{tab:matrixdimension} for the linear dimension $L$ of the corresponding parent cubic lattice. (b) TDOS at finite energies $\rho^{\rm PWB}_{\rm typ} (E)$ in PWB containing 2755 sites as a function of $W$. (c) Critical value  of disorder $W_{c}$ for the semimetal-to-metal quantum phase transition as a function of $(N^{\rm 2D}_s)^{-1}$, demonstrating its convergence to a finite value of $0.75 \pm 0.10$ in the thermodynamic limit $(N^{\rm 2D}_s)^{-1} \to 0$, which agrees well with the critical disorder strength for this transition obtained from the scaling of the subtracted average density of states at zero energy, see Figs.~\ref{fig:PBWeylADOS}(c) and~\ref{fig:PBWeylADOS}(f). (d) $\rho^{\rm PWB}_{\rm typ} (0)$ as a function of $W$ for a wide range of disorder strength on a PWB with $N^{\rm 2D}_s=2755$, showing Anderson localization at sufficiently large disorder of $W_c=20.00 \pm 1.00$ with its inset showing the semimetal-to-metal transition. The order parameter exponent ($\beta$) obtained from the scaling of the TDOS near (e) the semimetal-to-metal transition yielding $\beta=1.80 \pm 0.10$ and (f) Weyl metal-to-Anderson insulator transition yielding $\beta=1.60 \pm 0.15$. All the results are obtained by projecting the parent Weyl Hamiltonian on a clean cubic lattice with periodic boundary condition in each direction to the PWB and subsequently adding disorder therein, and from exact diagonalization. TDOS is computed after averaging the local density of states over $N_t=4$ sites within the PWB and $500$ independent disorder realizations.     
     }~\label{fig:WeylAnderson}
\end{figure}

\subsection{Typical density of states in Weyl systems}

So far we have used the (subtracted) ADOS at zero energy to underpin the Weyl semimetal-to-metal transition on cubic lattices and 2D PWBs. In this subsection, we compute TDOS at zero energy to identify this transition as well as the Weyl metal to Anderson insulator transition. It should be emphasized that the spurious zero energy states of the clean Weyl Hamiltonian, detailed in Appendix~\ref{append:zeromodes}, are extended plane wave states and thus continue to cause issues in identifying the semimetal-to-metal transition via the scaling of TDOS at zero energy. Additionally, when computing the TDOS we can not utilize the procedure of subtracting the zero mode contribution in the clean limit, implemented while computing ADOS. This is because, with increasing disorder strength, these spurious zero modes gradually localize, making their identification and removal nontrivial. To circumvent this issue and in order to reliably determine $W_{c}$ for the semimetal-to-metal transition from the scaling of TDOS at zero energy, we therefore utilize a modified procedure in which, upon diagonalizing the disordered Hamiltonian exactly, the eigenstates are sorted in ascending or descending order of the energy eigenvalues. For a Hamiltonian consisting of $2N$ eigenvalues, we then do not consider the $(N-1)$th and $N$th eigenstates near the band center in computation of the TDOS which amounts to removing the eigenstates corresponding to the zero modes prior to computation of the TDOS. Here, we exclusively focus on the disordered PWB and the results are shown in Fig.~\ref{fig:WeylAnderson}, which we discuss in the following  paragraph.

In Fig.~\ref{fig:WeylAnderson}(a), we show the dependence of the TDOS at zero energy $\rho^{\rm PWB}_{\rm typ}(0)$, computed following the procedure outlined above, on PWBs with varying total number of sites therein $N^{\rm 2D}_s$. The semimetal-to-metal transition is identified by the strength of disorder ($W_c$) at which $\rho^{\rm PWB}_{\rm typ}(0)$ becomes finite, which occurs at decreasing strength of disorder with increasing $N^{\rm 2D}_s$. In Fig.~\ref{fig:WeylAnderson}(b) we display the scaling of $\rho^{\rm PWB}_{\rm typ}(E)$ over a wide range of disorder strength, which closely mimics the scaling of ADOS with disorder on PWBs, shown in Fig.~\ref{fig:PBWeylADOS}(b) and~\ref{fig:PBWeylADOS}(e). In Fig.~\ref{fig:WeylAnderson}(c), we analyze the scaling of $W_c$ for the semimetal-to-metal transition, obtained from the scaling of $\rho^{\rm PWB}_{\rm typ}(0)$, which saturates to a value of $0.75 \pm 0.10$ in the thermodynamic limit as $(N^{\rm 2D}_s)^{-1} \to 0$, in reasonable agreement with the the value of $W_c=0.70 \pm 0.02$ obtained from the scaling of ADOS at zero energy, although these two values are strictly expected to be \emph{identical} as is the case on a cubic lattice~\cite{pix2015, Slager2018}.

In Fig.~\ref{fig:WeylAnderson}(d) we display the scaling of $\rho^{\rm PWB}_{\rm typ}(0)$ in the largest PWB with $N^{\rm 2D}_s=2755$ over a much broader range of disorder, which unfolds the projected Weyl metal to Anderson insulator localization transition at disorder strength $W_c=20.00 \pm 1.00$. From the scaling of the TDOS at zero-energy near the semimetal-to-metal and metal-to-insulator transitions, we extract the corresponding order parameter exponents, respectively, yielding $\beta=1.80 \pm 0.10$ as shown in Fig.~\ref{fig:WeylAnderson}(e) and $\beta=1.60 \pm 0.15$ as shown in Fig.~\ref{fig:WeylAnderson}(f). Values of these order parameter exponents on PWBs, obtained from the scaling of $\rho^{\rm PWB}_{\rm typ}(0)$, are sufficiently close to the ones previously reported for cubic lattice-based Weyl models~\cite{Slager2018}. Also notice that the values of the exponent  $\beta$ extracted near the semimetal-to-metal transition from the scaling of ADOS and TDOS differ, suggesting that the wavefunctions near this transition on PWBs exhibit multifractality, consistent with prior observations in cubic lattice-based simulations~\cite{pix2015, Slager2018}. A detailed numerical investigation of the multifractal nature of wavefunctions on disordered PWBs is left for future work.  The fact that the scaling and critical behavior of $\rho^{\rm PWB}_{\rm typ}(0)$ essentially match those on cubic lattices further provides compelling evidence that PWBs constitute genuine quantum holographic descendants of their parent cubic Weyl systems.

\section{Summary and discussion}~\label{sec:summary}

To summarize, here we study the effects of disorder on 2D PBs that are constructed as quantum holographic images from their parent 3D cubic crystals via the Schur decomposition (Fig.~\ref{fig:construction}). We focus on two prominent representatives for 3D disordered systems: the Anderson model and dirty Weyl semimetals. In three dimensions, while the former one only displays the Anderson metal-insulator transition at strong disorder, the Weyl semimetals subject to on-site random impurities exhibit two distinct quantum phase transitions. At first, they undergo a semimetal-to-metal transition at moderate disorder strength, followed by a second one, the traditional metal-insulator Anderson transition. The TDOS at zero-energy (corresponding to the band center in particle-hole symmetric models) serves as the bona fide order-parameter to identify both the transitions, while the ADOS at zero-energy emerges as an additional order parameter across the Weyl semimetal-to-metal transition. Each of these transitions is characterized by distinct critical exponents, which are known in the literature to high accuracies~\cite{Anderson1958, thouless1970, AALR1979, Wegner1979, Lee1981, Hikami1981, RevModPhys1985AT, Kirkpatrick1994, janssen1998, slevin1999, slevin2001, klopp2002, dobrosavljevic2003, Brndiar2006, Chakravarty2008, RevModPhys2008AT, slevin2010, Fradkin1986, Murakami2009, goswami2011, Imura2013, Herbut2014, roydassarma2014, Syzranov2015a, Syzranov2015b, Altland2015, pix2015, Syzranov2016, Pix2016a, roydassarma2016, bera2016, Gorbar2016, RoyJuricicDasSarma2016, RoyDasSarma2016PRB, Carpentier2016, Carpentier2017, Justin2017, Goswami2017a, Goswami2017b, Alavirad2017, Slager2017, WilsonRefael2018, Slager2018, Carpentier2018, Carpentier2019, Szabo2020, Gruzberg2024}.

In this work, we sought to harness the quantum phase diagrams of such disordered 3D systems and the associated quantum criticality (in terms of the critical exponents) on their projected 2D branes for which the effective Hamiltonian are constructed via the Schur decomposition from the parent cubic lattice-based Hamiltonian. Fascinatingly, we find that the quantum phase diagram of such 2D disordered hexagonal brane, embedded within the parent cubic lattice, not only manifests all the phases of the dirty 3D crystals and the quantum phase transitions among them, the numerically extracted critical exponents across various quantum phase transitions are also sufficiently close to the ones known in 3D systems. See also Table~\ref{tab:summary_results}. Altogether, these findings open a promising route to emulate disorder-driven quantum phenomena that are specific to 3D crystals on their geometric descendants in the form of 2D hexagonal branes.

We believe that the central findings of this work, as summarized above, are not unique to 2D branes of 3D parent crystals. For instance, we anticipate that analogous results would also manifest on 1D branes of the 3D cubic lattice aligned along its body diagonal. However, a numerical verification of this conjecture would require the inversion of even larger matrices ($H_{22}$) to construct the effective Hamiltonian for a 1D brane comprising a sufficiently large number of lattice sites, as detailed in Table~\ref{tab:matrixdimension}. This task is significantly more computationally demanding than the corresponding analysis for the 2D branes and unfortunately exceeds the capacity of our current numerical resources. Consequently, we must leave this claim at the level of a \emph{conjecture}, which we hope to address in future investigations. We also believe that such outcomes should be insensitive to the orientation of two- and one-dimensional PBs within the parent cubic lattice, which we intend to demonstrate in the future.

It is important to note that the quantum phase diagrams of dirty PBs and phase transitions therein are qualitatively, and, within numerical accuracy, quantitatively identical regardless of whether one first projects the clean Hamiltonian from three to two dimensions and subsequently introduces disorder, or directly projects the disordered 3D Hamiltonian. This equivalence is explicitly demonstrated in this work for the Weyl semimetal-to-metal transition; see Fig.~\ref{fig:PBWeylADOS} and Table~\ref{tab:summary_results}. Therefore, in the context of material realizations of quantum phases and phase transitions of parent 3D crystals on their 2D branes, experimental efforts can focus primarily on engineering the clean Hamiltonian for the 2D brane with the highest possible accuracy, this being the cornerstone of our proposal, as discussed in more details below, and subsequently introducing disorder therein. In practice, the stochastic averaging over multiple independent disorder realizations in metamaterial-based experiments can be accomplished by designing a single realization of a clean brane and then systematically modulating the disorder configurations therein.

\begin{figure}[t!]
\includegraphics[width=1.00\linewidth]{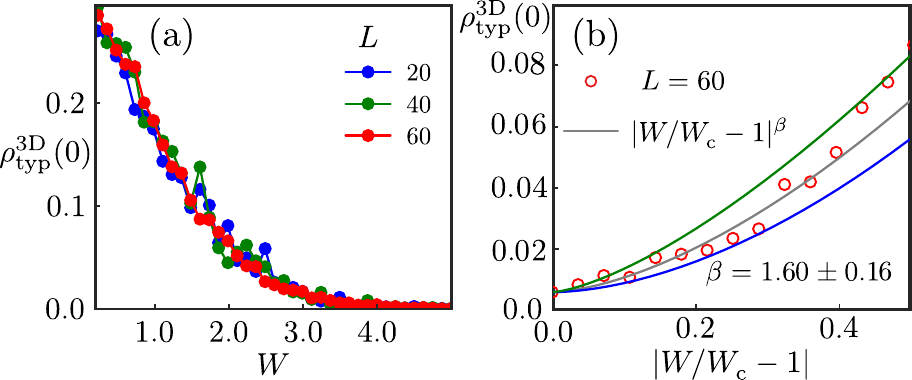}
     \caption{(a) Typical density of states (TDOS) at zero energy of a three-dimensional (3D) Anderson model $\rho^{\rm 3D}_{\rm typ}(0)$, computed using the kernel polynomial method on cubic lattices of varying linear dimensionality $L$ (see legend) in each direction. We impose periodic boundary condition in all directions. The critical disorder strength $W_c=3.10 \pm 0.10$ for the Anderson transition is identified from the vanishing of $\rho^{\rm 3D}_{\rm typ}(0)$. (b) Extraction of the order parameter exponent $\beta=1.60 \pm 0.16$ from the scaling of TDOS at zero energy near the Anderson transition, see Eq.~\eqref{eq:Andersonscaling}, which is in agreement with the results obtained from exact diagonalization on a $L=20$ cubic lattice and projected two-dimensional brane containing only $N^{\rm 2D}_s=2161$ sites therein. See Fig.~\ref{fig:BraneAnderson} for comparison. TDOS is computed after averaging the local density of states over $N_t=12$ sites within the cubic lattice and we compute $N_c=4096$ Chebyshev moments and average over 100 independent disorder realizations to extract $\rho^{\rm 3D}_{\rm typ}(0)$.     
     }~\label{fig:AndersonKPM}
\end{figure}

Our theoretical predictions for disordered 2D branes, which capture the quantum phase diagram of parent 3D dirty crystals, including the semimetal-to-metal and Anderson metal-insulator transitions, can be experimentally tested across a variety of metamaterial platforms that support wave dynamics, with photonic lattices being particularly promising in light of recent experiments. It is important to emphasize that these phenomena fundamentally stem from the wave-like nature of electrons, with their dynamics governed by the Schrödinger equation, in the presence of disorder. Consequently, any metamaterial system that emulates wave phenomena can, in principle, exhibit these effects~\cite{Anderson1985}. Among such platforms, photonic lattices stand out~\cite{Segev2013, Abdullaev1980, SajeevJohn1984, DeRaedt1989}, where disorder can be introduced by random modulations of the refractive index, and Anderson localization of light has already been observed experimentally in both two~\cite{Schwartz2007,wang2024observation} and one~\cite{Lahini2008,lahini2009observation,szameit2010wave} dimension. While photonic lattices inherently support only transverse localization of light, potentially limiting the realization of phenomena such as the Anderson transition, characteristic of three dimensions, this limitation remains an open question due to the unique propagation properties of photons~\cite{Skipetrov2016}. Our results thus chart a realistic path toward experimentally realizing the 3D Anderson transition and the semimetal-to-metal transition in suitably designed photonic lattices, as described by the effective hopping Hamiltonian in Eq.~\eqref{eq:PTB}, which is also supported by the recent advances in disorder engineering in photonic systems~\cite{yu2021engineered}.  Furthermore, recent experimental advances have demonstrated Anderson localization of elastic waves in 3D disordered media~\cite{Hu2008} and of light in biological media~\cite{Choi2018}. Altogether, these highly tunable metamaterial platforms can be systematically engineered to emulate the effective Hamiltonian of a 2D PB with disorder, thereby enabling the exploration of disorder-induced phenomena emerging from  the wave nature of electrons, light, and elastic waves.

Although the proposed metamaterial platforms present a promising avenue for the experimental realization of 3D Anderson and semimetal-to-metal transitions, guided by our general principle for construction of the effective Hamiltonian on 2D projected branes, there are inherent practical limitations to such implementations. Specifically, while the tight-binding Hamiltonian for the parent 3D cubic lattice (as in the Anderson and Weyl models) involves only nearest-neighbor hopping process [see Eq.~\eqref{eq:parent}], the effective Hamiltonian for the 2D brane features long-range hopping across the entire system, which becomes `infinitely' long-ranged in the thermodynamic limit due to $H^{-1}_{22}$ in the second term in Eq.~\eqref{eq:PTB}. Engineering such infinitely long-range hopping in metamaterial platforms poses significant technical challenges.

\begin{figure*}[t!]
\includegraphics[width=1.00\linewidth]{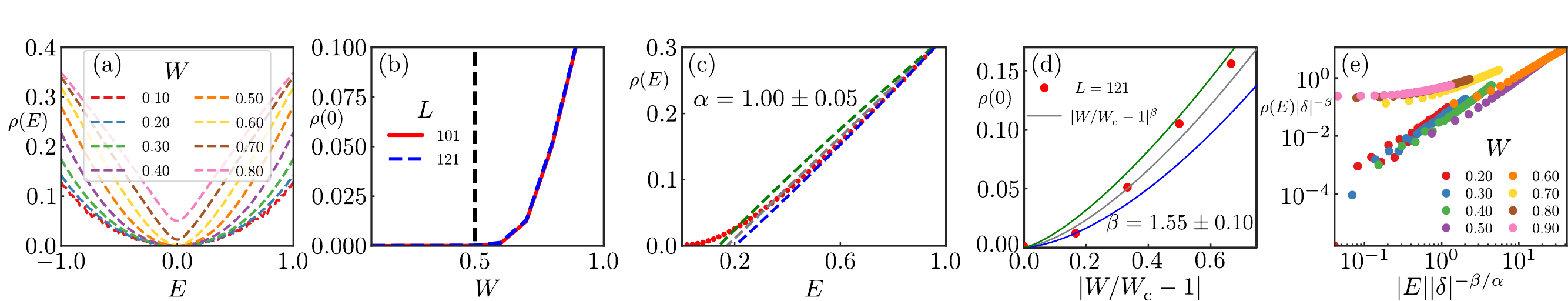}
     \caption{(a) Average density of states (ADOS) for a three-dimensional Weyl semimetal, see Eq.~\eqref{eq:ADOS}, as a function of disorder strength $W$ (see legend) on a cubic lattice of a linear dimension $L=121$ and periodic boundary condition in each direction, obtained after averaging over 100 independent disorder configurations. Results are obtained by employing the kernel polynomial method with $N_{c}=2024$ Chebyshev moments. Stochastic trace is computed from 12 random unimodular vectors. (b) Comparison of ADOS at zero energy as a function of disorder strength in $L=101$ and $L=121$ cubic lattices. The almost perfectly overlapping curves in these two systems indicate that the finite size effects are minimal and the semimetal-to-metal transition occurs at critical disorder strength $W_{c}=0.50 \pm 0.05$ (marked by the vertical black dashed line). Compare with Fig.~\ref{fig:DOSWeyl2D3D}(c) yielding $W_c=0.49 \pm 0.05$ from the scaling of the subtracted ADOS at zero energy in a much smaller cubic lattice of $L=19$. Extraction of (c) the ADOS scaling exponent, yielding $\alpha=1.00 \pm 0.05$ and (d) order parameter exponent, yielding $\beta=1.55 \pm 0.10$ in $L=121$ system. (e) Collapse of data onto three branches from the comparison of $|E| \delta^{-\beta/\alpha}$ and $\rho(E) \delta^{-\beta}$. For comparison of results from exact diagonalization on a cubic lattice of $L=19$ and projected Weyl branes, see Fig.~\ref{fig:exponents}. 
     }~\label{fig:WeylKPM}
\end{figure*}

Nevertheless, it is noteworthy that the critical exponents associated with the Anderson and semimetal-to-metal quantum phase transitions on the 2D projected branes are in close quantitative agreement with those of the parent 3D systems. This striking correspondence strongly indicates that the presence of long-range hopping on the 2D brane effectively restores a dimensionality of three therein. Conversely, omitting the second term in Eq.~\eqref{eq:PTB} (setting $\kappa=0$ therein) reduces the model to a conventional 2D lattice with only nearest-neighbor hopping, devoid of both Anderson and semimetal-to-metal transitions. These contrasting observations motivate a following  proposal: by systematically reducing the range of hopping on the 2D brane, achievable in Eq.~\eqref{eq:PTB} by introducing a real-space Heaviside $\Theta$ function in its second term, the effective dimensionality of the system interpolates smoothly between three and two. Given that the lower critical (LC) dimension for both Anderson and semimetal-to-metal transitions is $d_{\rm LC}=2$, we argue that as long as the hopping on the 2D brane extends beyond strictly nearest neighbors, both transitions therein remain possible. However, the unambiguous manifestation of these transitions necessitates that the hopping retains at least a moderately long-ranged character. Furthermore, as the hopping range is gradually reduced, we expect the critical exponents ($\alpha$ and $\beta$) to vary \emph{continuously}, ultimately converging to values characteristic of a system with only nearest-neighbor hopping, where no transition occurs. A comprehensive theoretical demonstration of these predictions, via extensive numerical analysis, is a subject we leave for future investigation.

Crucially, the requirement for only moderately long-range hopping renders the experimental realization of these 2D PB Hamiltonian feasible with current metamaterial design capabilities. Thus, our theoretical predictions for the observation of robust semimetallic, metallic, and insulating phases, as well as the quantum phase transitions between them on 2D holographic projected disordered branes (see Fig.~\ref{fig:Phasediag}) hold significant promise for their experimental realizations in the near future.

\acknowledgments 
V.J.\ acknowledges the support of the Swedish Research Council Grant No.\ VR 2019-04735 and Fondecyt (Chile) Grant No.~1230933. Nordita is supported in part by NordForsk. The computations were enabled by resources provided by the National Academic Infrastructure for Supercomputing in Sweden (NAISS), partially funded by the Swedish Research Council through grant agreement No.\ 2022-06725. B.R.\ was supported by NSF CAREER Grant No.~DMR-2238679 (USA) and thanks ANRF, India, for support through the Vajra scheme VJR/2022/000022. We are in debt to Christopher A.\ Leong for helps with finalizing the figures and critical reading of the manuscript. 
\appendix

\section{Quantization and zero modes}~\label{append:zeromodes}

In the main manuscript, we pointed out that for the prototypical model of Weyl semimetal, see Eq.~\eqref{eqn:fullH}, utilized in this work, the exact diagonalization under periodic boundary conditions in a cubic lattice yields two and sixteen zero-energy states for a system of linear dimension size $L=2M+1$ and $L=2 M$, respectively, where $M$ is an integer ($\mathbb{Z}$). In this Appendix, we unfold the origin of such zero-energy modes. For a cubic lattice of linear dimension $L$ in each direction with unit lattice spacing ($a=1$), the momenta take on discrete quantized values, given by $k_{j}=2\pi q_{j}/L$, where $q_{j}\in \mathbb{Z}$. In order to impose periodic boundary conditions, we require $\exp[ik_{j}(L+1)]=\exp[ik_{j}]$, implying that $\exp[ik_{j}L]=1$. The energy spectrum in such a finite geometry of the cubic lattice is given by  
\begin{equation}
    E(q_{x},q_{y},q_{z})=\pm t \left[ \sum_{j=x,y,z}\sin^2 \left( 2\pi q_{j}/L \right) \right]^{1/2}. 
\end{equation}
For a system where $L$ is odd, we then find that the choice $q_x=q_y=q_z=0$ yields exactly two zero-energy states. However, when $L$ is even, both $q_x=q_y=q_z=0$ and $q_x=q_y=q_z=L/2$, and their combinations yield total sixteen zero-energy states. To minimize the impact of such zero-energy modes in the computation of the ADOS at and near zero-energy, we work with cubic lattices with odd $L$. Similar conclusion holds for a square lattice and we thus consider square lattices for which $L$ is odd.

\section{Results from kernel polynomial method}~\label{append:KPM}

To establish the reliability of the exact diagonalization method in accurately capturing various phases, identifying critical disorder strengths for different phase transitions, and extracting the corresponding critical exponents in disordered systems, we systematically compare its results to those obtained via the KPM. While our primary analysis on 2D PBs relies on exact diagonalization (in smaller systems), here we benchmark this approach against KPM for cubic lattices, which allows access to much larger system sizes.  We begin this comparison for the Anderson model. We use the KPM to compute the TDOS at zero energy on cubic lattices with linear dimensions $L=20,40$, and $60$ and with periodic boundary conditions in each direction, from which we identify the critical disorder strength for the Anderson transition to be $W_c=3.10 \pm 0.10$ in $L=60$ system, as shown in Fig.~\ref{fig:AndersonKPM}(a), which is, however, reasonably insensitive to the system size ($L$) and agrees well with the value obtained from exact diagonalization in the $L=20$ system [see Fig.\ref{fig:BraneAnderson}(a)].  The order parameter exponent, obtained from the scaling of TDOS at zero energy near the Anderson transition is found to be $\beta=1.60 \pm 0.16$ in a $L=60$ cubic lattice as shown in Fig.~\ref{fig:AndersonKPM}(b), which is also close to the one we found from exact diagonalization in $L=20$ cubic lattice, displayed in Fig.~\ref{fig:BraneAnderson}(b).

As discussed in the main part of the manuscript, a computational bottleneck arises in the study of disordered PWBs due to the necessity of accurately computing the inverse of the matrix $H_{22}$ in Eq.\eqref{eq:PTB}. This precludes the use of sparse matrix methods such as Lanczos~\cite{lanczos1950} and KPM~\cite{kpm}, which are typically employed to efficiently access system sizes approaching the thermodynamic limit and thereby substantially reduce the finite-size effects. Consequently, in this work, we rely on exact diagonalization and scaling analysis as a function of system size to determine $W_c$ in the thermodynamic limit for the semimetal-to-metal transition in PWBs, and we compare these results with those obtained for cubic lattices using the same methodology. In this appendix, we further utilize KPM to determine $W_{c}$ for the prototypical cubic lattice-based model Hamiltonian for 3D Weyl semimetal in terms of nearest-neighbor hopping amplitudes from Eq.~\eqref{eqn:fullH} in sufficiently large systems and compare the results obtained from the exact diagonalization in smaller cubic systems.

The KPM enables expansion of the property of interest, in our case the ADOS, in terms of the Chebyshev polynomial to an order $N_{c}$. This efficient process allows sparse matrix methods to be leveraged and extremely large system sizes to be accessed. We utilize KPM to analyze the effects of disorder for the Hamiltonian from Eq.~\eqref{eqn:fullH}, considering $L=101$ and $L=121$ cubic lattices. We note that the effect of the zero modes analyzed in the previous Appendix is minimized in such large systems, as their contribution to the ADOS is given by Eq.~\eqref{eq:ados3} scales as $1/N$, where $N$ is the number of lattice sites in the system. For the model considered, $N = 2L^3$ for a cubic lattice. Thus, for $L > 100$, the contribution from the zero-energy modes is negligible when utilizing KPM.

\begin{figure}[t!]
\includegraphics[width=1.00\linewidth]{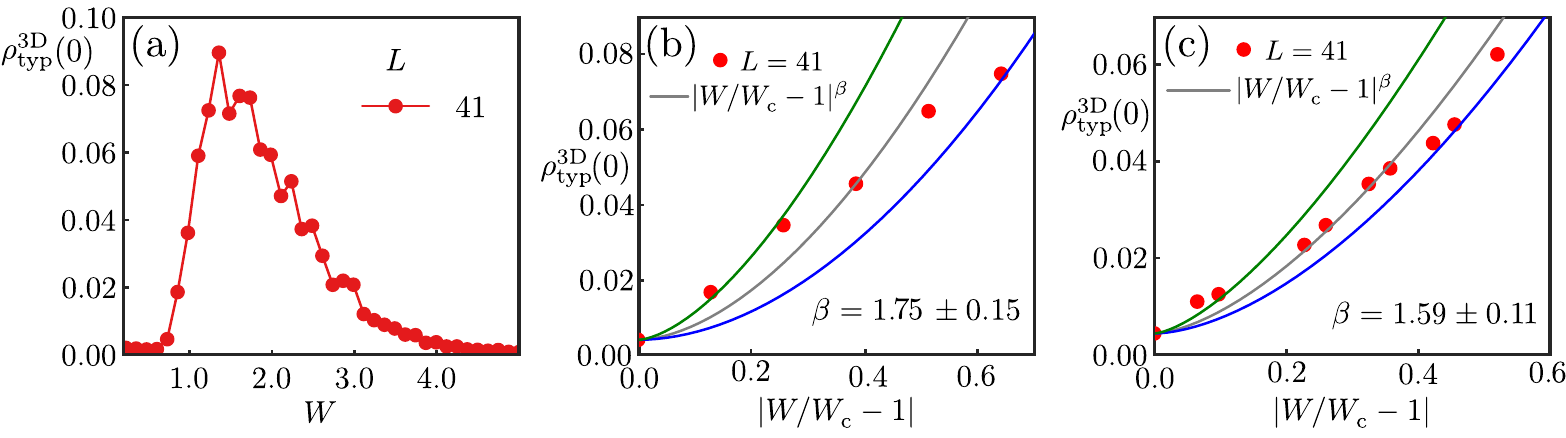}
     \caption{(a) Variation of the typical density of states (TDOS) at zero energy $\rho^{\rm 3D}_{\rm typ}(0)$ for a three-dimensional Weyl system as a function of disorder strength ($W$), obtained from kernel polynomial method on a cubic lattice of linear dimension $L=41$ and periodic boundary condition in each direction. It shows a semimetal-to-metal transition at weak disorder $W_c=0.70 \pm 0.10$ and a metal-Anderson insulator transition at large values of disorder $W_c=3.85 \pm 0.15$. Extraction of the order-parameter exponent ($\beta$) from the power-law scaling of the form $\rho^{\rm 3D}_{\rm typ}(0)\sim |W/W_{c}-1|^{\beta}$ near the (b) semimetal-metal transition yielding $\beta=1.75 \pm 0.15$ and (c) metal-Anderson insulator transition, yielding $\beta=1.59 \pm 0.11$. TDOS is computed after averaging the local density of states on four sites ($N_t=4$) and computing $N_c=4096$ Chebyshev moments over 200 independent disorder realizations.   
     }~\label{fig:WeylKPMTDOS}
\end{figure}

The ADOS from the KPM is obtained by fixing the number of Chebyshev moments to be $N_{c}=2024$ and averaging over 100 disorder configurations. The stochastic trace is computed from 12 unimodular random vectors. The results are shown in Fig.~\ref{fig:WeylKPM}. The scaling of the average DOS $\rho(E)$ over a wide range of disorder strength in a $L=121$ cubic lattice is shown in Fig.~\ref{fig:WeylKPM}(a), which qualitatively agrees with the one we found from the scaling of the subtracted ADOS using exact diagonalization. The scaling of the average DOS at zero-energy is almost identical in $L=101$ and $L=121$ cubic lattices, as shown in Fig.~\ref{fig:WeylKPM}(b). The critical disorder strength for the Weyl semimetal-to-metal transition, estimated from the KPM, is $W_{c}= 0.50 \pm 0.05$, in very good agreement with the value determined in the main part of the manuscript on a smaller cubic system through a scaling analysis as a function of system size, yielding $W_{c} = 0.49 \pm 0.05$. From the scaling of ADOS, we find the DOS scaling exponent $\alpha=1.00 \pm 0.05$ [Fig.~\ref{fig:WeylKPM}(c)] and order parameter exponent $\beta=1.55 \pm 0.10$ [Fig.~\ref{fig:WeylKPM}(d)] in $L=121$ cubic lattices. Finally, for $\alpha=1.00$ and $\beta=1.55$, we find the ADOS data collapse onto three branches when we compare $|E| |\delta|^{-\beta/\alpha}$ and $\rho(E) \delta^{-\beta}$, as shown in [Fig.~\ref{fig:WeylKPM}(e)].

Finally, we employ the KPM to compute the TDOS at zero energy $\rho_{\rm typ}(0)$ for the Weyl Hamiltonian in a $L=41$ cubic lattice by introducing independent and random potential disorder on each site therein. Here, we compute $\rho_{\rm typ}(0)$ with $N_c=4096$ Chebyshev moments and after averaging over 200 independent disorder realizations. We note that $\rho_{\rm typ}(0)$ is computed from the disorder-averaged local DOS over four sites with $N_t=4$ in Eq.~\eqref{eq:tdos}. The results are shown in Fig.~\ref{fig:WeylKPMTDOS}. As expected, the TDOS at zero energy remains pinned to zero in the Weyl semimetal phase in the weak-disorder regime, and becomes finite only in the metallic phase. The critical disorder strength for the semimetal-to-metal transition obtained from the scaling of $\rho_{\rm typ}(0)$ is $W_c=0.70 \pm 0.10$, in agreement with the value of $W_c$ for this transition obtained from the scaling of the ADOS at zero energy from KPM in large cubic lattices (Fig.~\ref{fig:WeylKPM}) and from the scaling of the subtracted ADOS at zero energy from exact diagonalization in smaller cubic lattices, when extrapolated to the thermodynamic limit (Fig.~\ref{fig:PBWeylADOS}). The order-parameter exponent ($\beta$) obtained from the scaling form $\rho_{\rm typ}(0) \sim (W/W_c-1)^\beta$ in the metallic phase is $\beta=1.75 \pm 0.15$, also in agreement with the one obtained from a similar scaling in a smaller cubic lattice with exact diagonalization. We note that with increasing strength of disorder, $\rho_{\rm typ}(0)$ again becomes zero at stronger disorder $W_c=3.85 \pm 0.15$, indicating the Anderson metal-insulator transition, which agrees with the results on smaller cubic lattices  with exact diagonalization when extended to the thermodynamic limit. The order-parameter exponent for the Anderson transition is $\beta=1.59 \pm 0.11$, which also in good agreement with the result we previously found from exact diagonalization in smaller cubic lattices.

\bibliography{Ref_DirtyPB_Final}

\end{document}